\documentclass{emulateapj}
\usepackage{amsmath}
\usepackage{amssymb}
\usepackage{natbib}
\shorttitle{Beamforming Errors in MWA Antenna Tiles}
\shortauthors{Neben et al.}

\begin{document}


\title{Beamforming Errors in Murchison Widefield Array Phased Array Antennas and their effects on Epoch of Reionization Science}

\author{Abraham~R.~Neben$^{1}$,
Jacqueline~N.~Hewitt$^{1}$,
Richard~F.~Bradley$^{2,3}$,
Joshua~S.~Dillon$^{1,4}$,
G.~Bernardi$^{12,20}$,
J.~D.~Bowman$^{5}$, 
F.~Briggs$^{6}$,
R.~J.~Cappallo$^{10}$, 
B.~E.~Corey$^{10}$,
A.~A.~Deshpande$^{11}$, 
R.~Goeke$^{1}$,
L.~J.~Greenhill$^{9}$,
B.~J.~Hazelton$^{16}$, 
M.~Johnston-Hollitt$^{18}$,
D.~L.~Kaplan$^{17}$, 
C.~J.~Lonsdale$^{10}$, 
S.~R.~McWhirter$^{10}$,
D.~A.~Mitchell$^{7,19}$, 
M.~F.~Morales$^{16}$, 
E.~Morgan$^{1}$, 
D.~Oberoi$^{13}$, 
S.~M.~Ord$^{8,19}$,
T.~Prabu$^{11}$, 
N.~Udaya~Shankar$^{11}$, 
K.~S.~Srivani$^{11}$, 
R.~Subrahmanyan$^{11,19}$, 
S.~J.~Tingay$^{8,19}$, 
R.~B.~Wayth$^{8,19}$, 
R.~L.~Webster$^{14,19}$, 
A.~Williams$^{8}$, 
C.~L.~Williams$^{1}$}

\altaffiltext{1}{Kavli Institute for Astrophysics and Space Research, Massachusetts Institute of Technology, Cambridge, MA 02139, USA}
\altaffiltext{2}{Dept. of Electrical and Computer Engineering, University of Virginia, Charlottesville, VA, 22904}
\altaffiltext{3}{National Radio Astronomy Obs., Charlottesville, VA}
\altaffiltext{4}{Dept. of Astronomy and Berkeley Center for Cosmological Physics, Berkeley, CA, USA}
\altaffiltext{5}{School of Earth and Space Exploration, Arizona State University, Tempe, AZ 85287, USA}
\altaffiltext{6}{Research School of Astronomy and Astrophysics, Australian National University, Canberra, ACT 2611, Australia}
\altaffiltext{7}{CSIRO Astronomy and Space Science (CASS), PO Box 76, Epping, NSW 1710, Australia}
\altaffiltext{8}{International Centre for Radio Astronomy Research, Curtin University, Bentley, WA 6102, Australia}
\altaffiltext{9}{Harvard-Smithsonian Center for Astrophysics, Cambridge, MA 02138, USA}
\altaffiltext{10}{MIT Haystack Observatory, Westford, MA 01886, USA}
\altaffiltext{11}{Raman Research Institute, Bangalore 560080, India}
\altaffiltext{12}{Square Kilometre Array South Africa (SKA SA), Cape Town 7405, South Africa}
\altaffiltext{13}{National Centre for Radio Astrophysics, Tata Institute for Fundamental Research, Pune 411007, India}
\altaffiltext{14}{School of Physics, The University of Melbourne, Parkville, VIC 3010, Australia}
\altaffiltext{15}{Sydney Institute for Astronomy, School of Physics, The University of Sydney, NSW 2006, Australia}
\altaffiltext{16}{Department of Physics, University of Washington, Seattle, WA 98195, USA}
\altaffiltext{17}{Department of Physics, University of Wisconsin--Milwaukee, Milwaukee, WI 53201, USA}
\altaffiltext{18}{School of Chemical \& Physical Sciences, Victoria University of Wellington, Wellington 6140, New Zealand}
\altaffiltext{19}{ARC Centre of Excellence for All-sky Astrophysics (CAASTRO)}
\altaffiltext{20}{Department of Physics and Electronics, Rhodes University, PO Box 94, Grahamstown, 6140, South Africa}








\begin{abstract}
Accurate antenna beam models are critical for radio observations aiming to isolate the redshifted 21\,cm spectral line emission from the Dark Ages and the Epoch of Reionization and unlock the scientific potential of 21\,cm cosmology. Past work has focused on characterizing mean antenna beam models using either satellite signals or astronomical sources as calibrators, but antenna-to-antenna variation due to imperfect instrumentation has remained unexplored. We characterize this variation for the Murchison Widefield Array (MWA) through laboratory measurements and simulations, finding typical deviations of order $\pm10-20\%$ near the edges of the main lobe and in the sidelobes. We consider the ramifications of these results for image- and power spectrum-based science. In particular, we simulate visibilities measured by a 100\,m baseline and find that using an otherwise perfect foreground model, unmodeled beamforming errors severely limit foreground subtraction accuracy within the region of Fourier space contaminated by foreground emission (the ``wedge''). This region likely contains much of the cosmological signal, and accessing it will require measurement of per-antenna beam patterns. However, unmodeled beamforming errors do not contaminate the Fourier space region expected to be free of foreground contamination (the ``EOR window''), showing that foreground avoidance remains a viable strategy. 
\end{abstract}





\section{Introduction}

Efforts to observe the formation of the first galaxies during the Dark Ages and the subsequent Epoch of Reionization (EOR) are at the frontier of observational cosmology. Tomographic maps of neutral Hydrogen in the Intergalactic Medium at these redshifts, where the majority of the observable comoving volume of the Universe resides, will shed light on questions ranging from astrophysics and cosmology to particle physics (see \citet{FurlanettoReview, miguelreview, PritchardLoebReview, aviBook, zaroubi} for reviews). The extreme brightness temperature sensitivity needed to isolate this faint signal in the presence of bright galactic and extragalactic radio emission (foregrounds) and detector noise necessitates thousand-hour integrations and hundreds of antenna elements \citep[e.g.][]{parsons12b, beardsley13, nithya13, PoberNextGen}. This quest is highlighting characterization of antenna beam patterns, or primary beams, as crucial for high dynamic range calibration and foreground subtraction \citep[][and Pober et al. (in prep)]{MoorePolarization, jacobs2013,nithya15}.

Two types of antenna mismodeling are relevant: (1) mismodeling of the mean antenna beam pattern; and (2) neglect of antenna-to-antenna variation. Both limit calibration and foreground subtraction fidelity in ways ranging from the obvious effect of subtracting sidelobe sources with the wrong apparent fluxes to the uncertain manner in which beam-related calibration errors average down with time. Indeed, modeling of antenna-to-antenna variation was long suspected to be critical for 21\,cm observatories, and early analysis pipeline development focused on incorporating knowledge of per-antenna beams in data reduction \citep{moralesandmatejek,fhd}, or even fitting for them in real time \citep{mwarts}.

The Murchison Widefield Array (MWA) \citep{lonsdale09,tingay13,mwascience} is now operating along with other instruments such as the Precision Array for Probing the Epoch of Reionization (PAPER) \citep{paperinstrument, parsons14} and the LOw Frequency Array (LOFAR) \citep{lofar}. Analysis of the data from these arrays is bringing new urgency to the question of antenna beam patterns. Source-based methods have long been used to constrain the mean antenna beam using interferometer cross-correlations (visibilities) \citep[e.g.,][]{pober12, lofar,colegatesutinjo}. More recently, working towards \textit{in-situ} measurements of per-antenna beams both for the MWA and for the developing next generation Hydrogen Epoch of Reionization Array (HERA) \citep{PoberNextGen,Whitepaper5}, \citet{NebenOrbcommPaper} present a beam measurement system using the ORBCOMM satellite constellation, an idea also explored by \citep{zheng14}. Development of a  drone equipped with a radio transmitter is also underway for the same application \citep{drone1,drone2}.

\begin{figure}
\includegraphics[width=8.38cm]{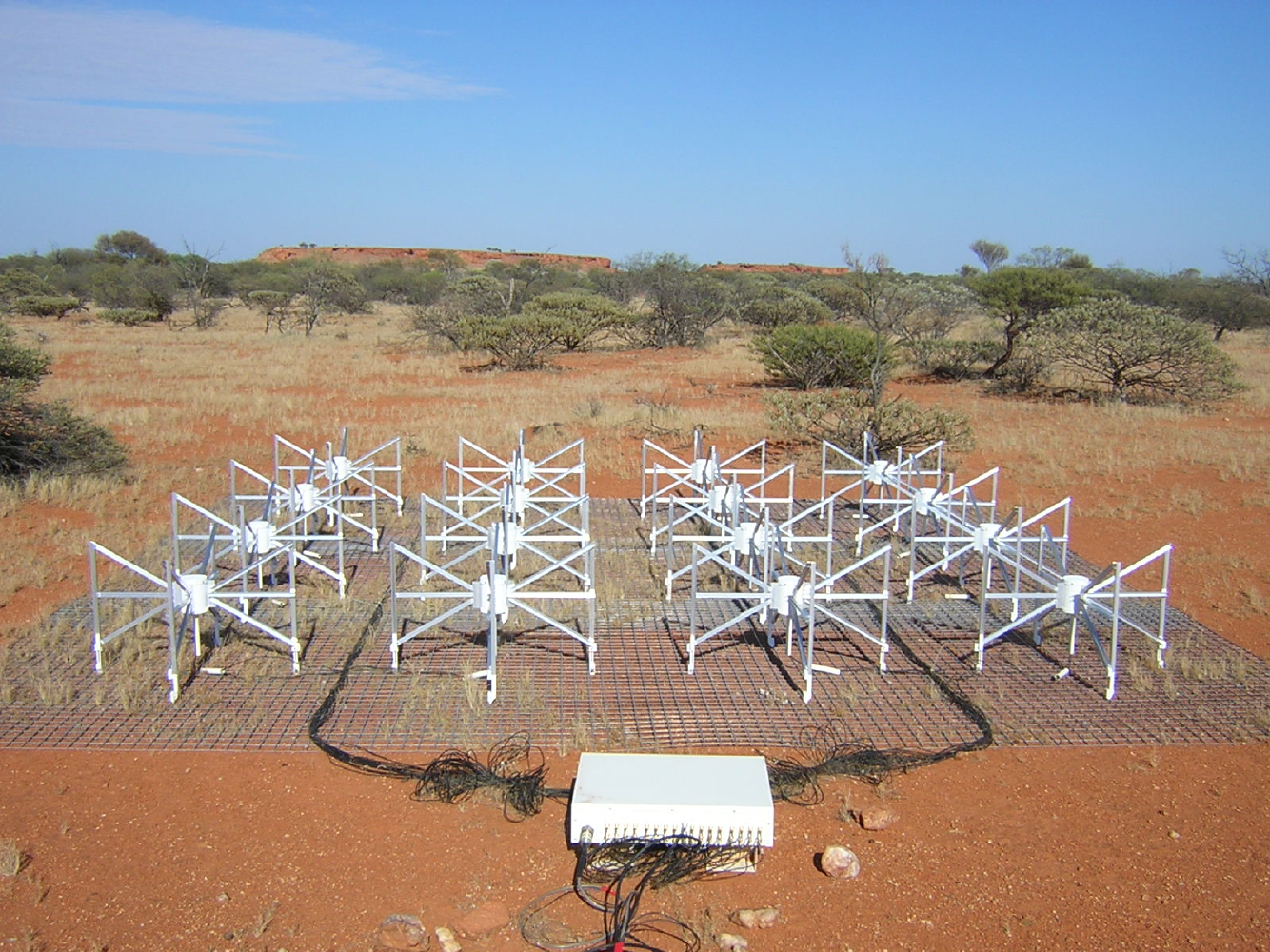}
\caption{One of the 128 deployed MWA tiles in the Murchison Radio Observatory, Western Australia.}
\label{fig:tilephoto}
\end{figure}

As the MWA uses $4\times4$ phased arrays of bowtie dipoles (hereafter MWA tiles) as its fundamental antenna elements, it is more prone to antenna-to-antenna beam variation than experiments with simpler antenna elements. PAPER has opted for simpler dipole-style elements at the expense of 24\,dB less zenith gain and increased risk of contamination by RFI and galactic emission near the horizon \citep{nithya15}. The cost of the MWA's larger per-element collecting area is sensitivity to group delay and gain matching errors which disrupt the coherent addition of dipole signals\footnote{For instance, if two -20\,dB reflections create a signal which adds $\pi/2$ out of phase with the main signal, a phase error of  $\sim1$deg is created, equivalent to a delay of 20\,ps at 150\,MHz}. LOFAR has similarly opted for phased array antennas and is developing direction-dependent calibration techniques to counter these systematics \citep{lofareorpaper}, and the issue is of particular import for the low frequency Square Kilometer Array (SKA-Low) \citep{ska1,ska2,ska3} whose design relies heavily on beamforming. Unfortunately, adding extra parameters to the calibration model tends to increase noise and risks cosmological signal loss \citep[e.g.,][]{gmrtsignalloss}.

As a first step towards understanding the magnitude of these effects to guide development of solutions like satellite- and drone-based beam calibration schemes, we focus in this paper on characterizing these beamforming errors in MWA tile beam patterns and begin to study their effects in a 21\,cm power spectrum analysis. In Section 2 we discuss laboratory measurements of beamforming errors and other systematics affecting the MWA tile, and compile a budget of beamforming errors. In Section 3 we study the effects of these errors on mean and standard deviation beam patterns using simulations, and consider the implications for EOR power spectrum measurements in Section 5. We discuss our results in Section 6. In order to put these beamforming errors into context and understand their origin and the trade-offs made in designing the MWA tile, we elaborate in Appendix A on the summary of the MWA tile presented by \citet{tingay13}.

\section{Laboratory Measurements of Beamforming Errors}
\label{sec:measurements}

\subsection{Overview of Beamforming in the MWA}

 \begin{figure*}[t]
 \centering
\includegraphics[height=2.5in]{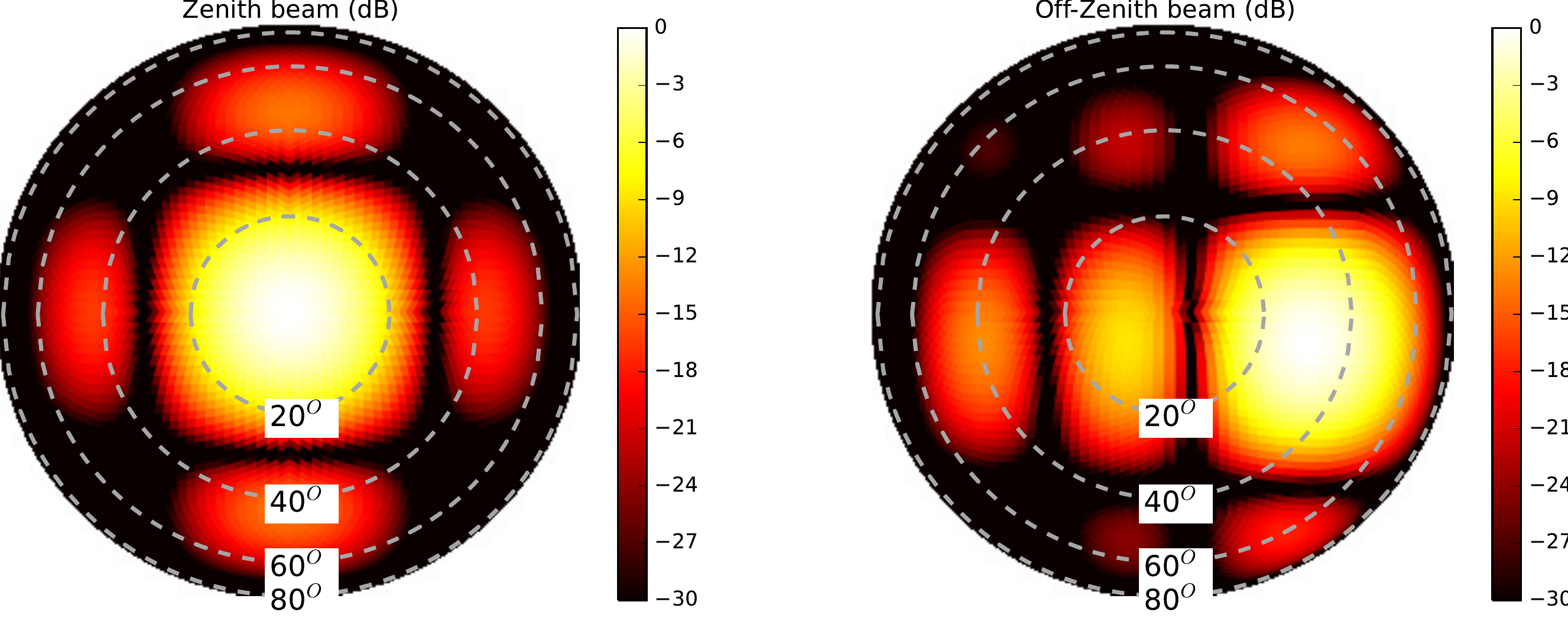}
\caption{Ideal (no beamforming errors) beams for a zenith pointing (left) and a representative off-zenith pointing (right) shown in sine-projection in units of dB at 150\,MHz. The off-zenith beam is pointed at $(\theta,\phi) = (53^\circ,101^\circ)$. }
\label{fig:idealbeams}
\end{figure*}

The Murchison Widefield Array consists of 128 antenna elements positioned in a centrally-concentrated, quasi-random distribution over a radius of 1.5\,km. Each antenna element (MWA tile) is a 4$\times$4 grid of dual-polarization bowtie dipoles with center-to-center spacing of 1.1\,m (half-wavelength at 136\,MHz) centered on a 5\,m$ \times $5\,m wire mesh ground screen (Figure \ref{fig:tilephoto}). The signals from the 16 antennas (each with a dual-polarization LNA) are summed in an analog beamformer with selectable delay lines, capable of applying phase gradients across the grid of dipoles to steer a beam of width full-width-at-half-max $25^\circ/(\nu/150\text{\,MHz})$ to elevations as low as $30^\circ$. We characterize the beamformer paths for delay bits 00000 (0\,ns) and 11111 (13\,ns); the actual EOR delays corresponding to elevations above $60^\circ$ are typically 5\,ns or smaller, and thus, in between these two cases. Figure \ref{fig:idealbeams} shows the zenith beam as well as a representative off-zenith beam. The first field tests on an early version of the MWA tile were presented by \citet{bowman07}, followed up by anechoic chamber measurements \citep{williamsthesis2012} and satellite-based measurements  \citep{NebenOrbcommPaper}.

\begin{figure*}[t]
\centering
\includegraphics[width=6in]{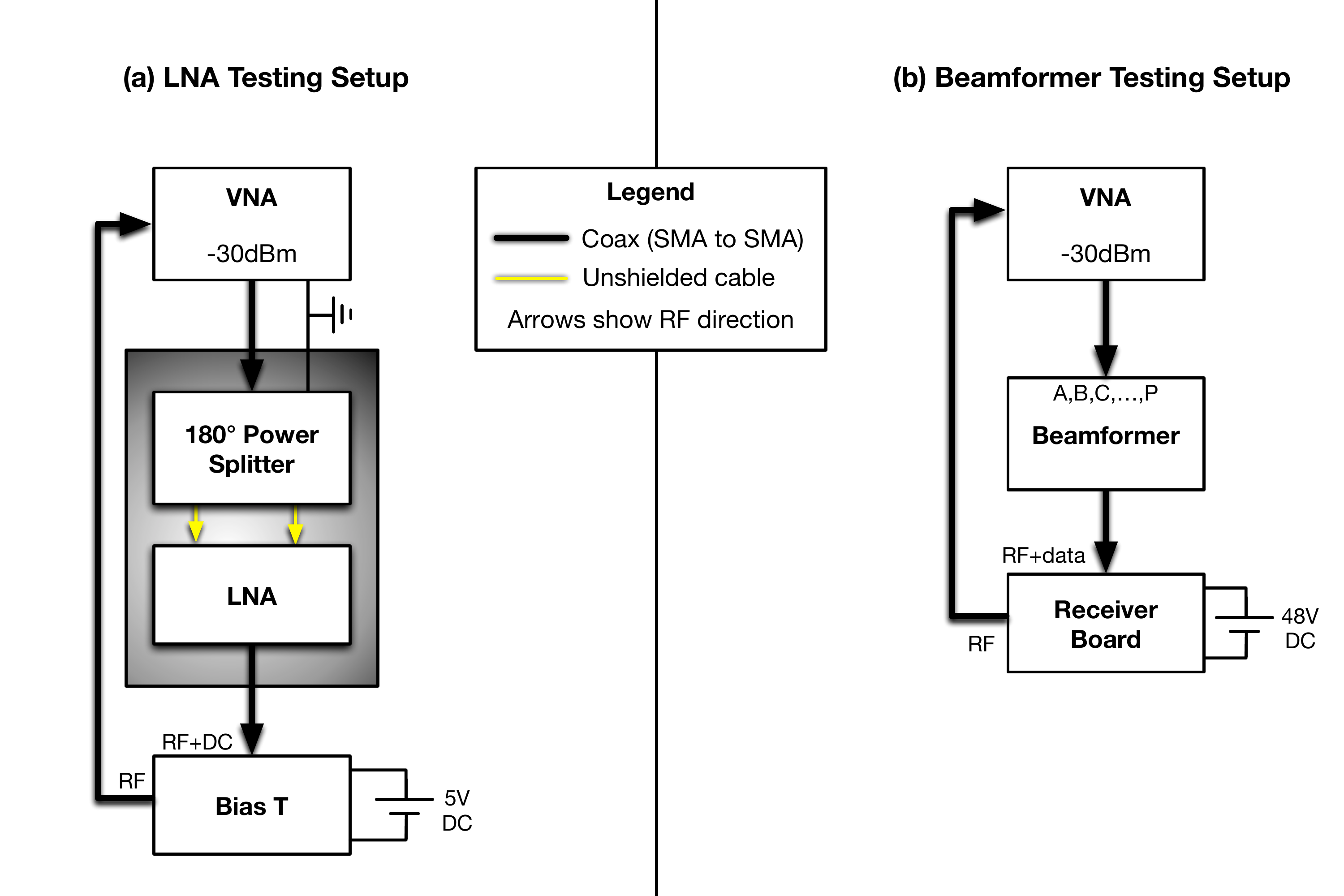}
\caption{Diagram showing our LNA and beamformer testing setups (Sec. \ref{sec:lnameasurements}). Note that LNA measurements are conducted above a ground plate to mitigate the effects of exposed antenna leads.}
\label{fig:experimentalsetup}
\end{figure*}

\begin{figure}[b]
\centering
\includegraphics[width=3in]{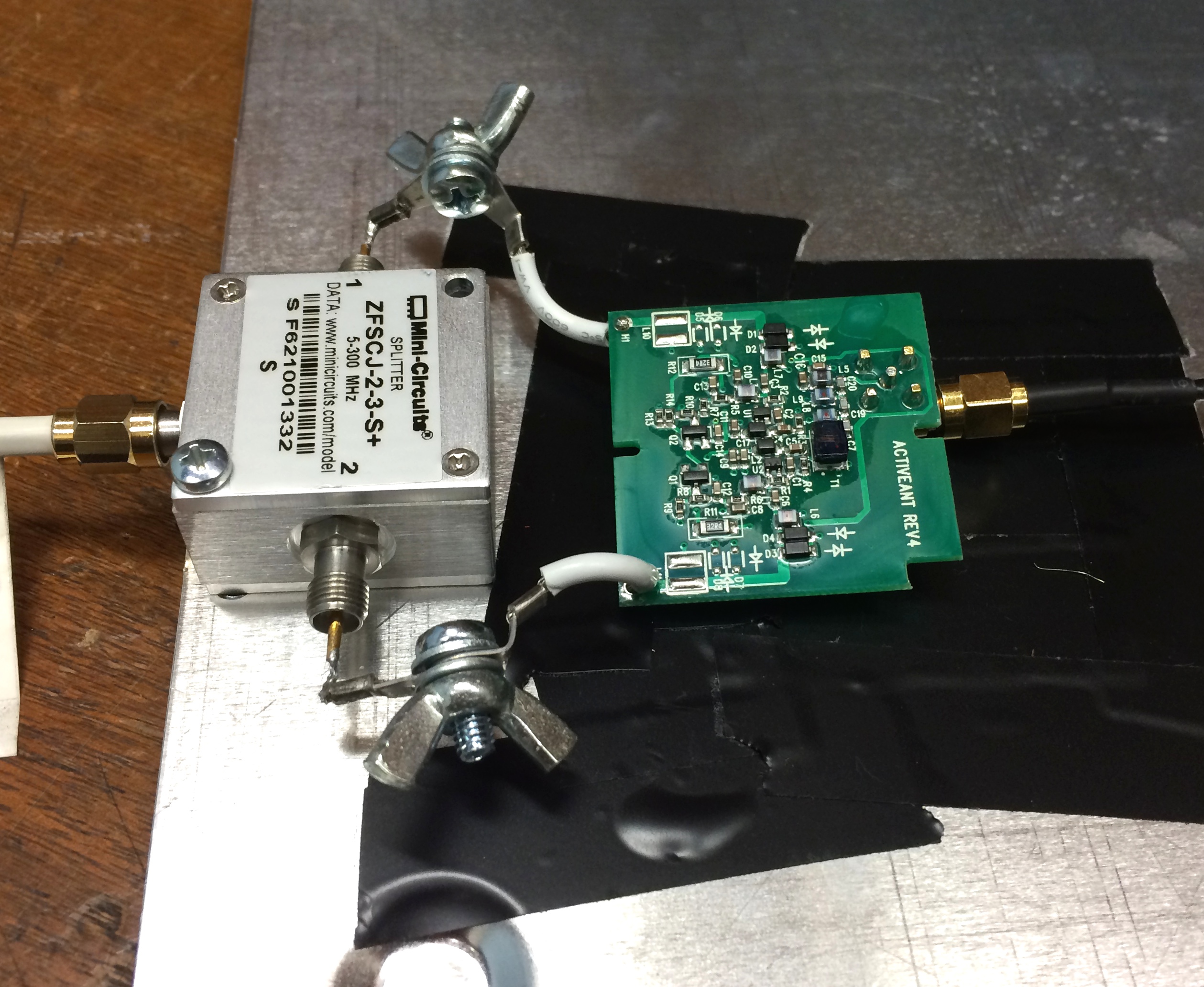}
\caption{Photograph of the LNA ground plate setup depicted in Figure \ref{fig:experimentalsetup} and described in Sec. \ref{sec:lnameasurements}.}
\label{fig:newlnasetup}
\end{figure}

\begin{figure*}[h]
\centering
\includegraphics[width=6in]{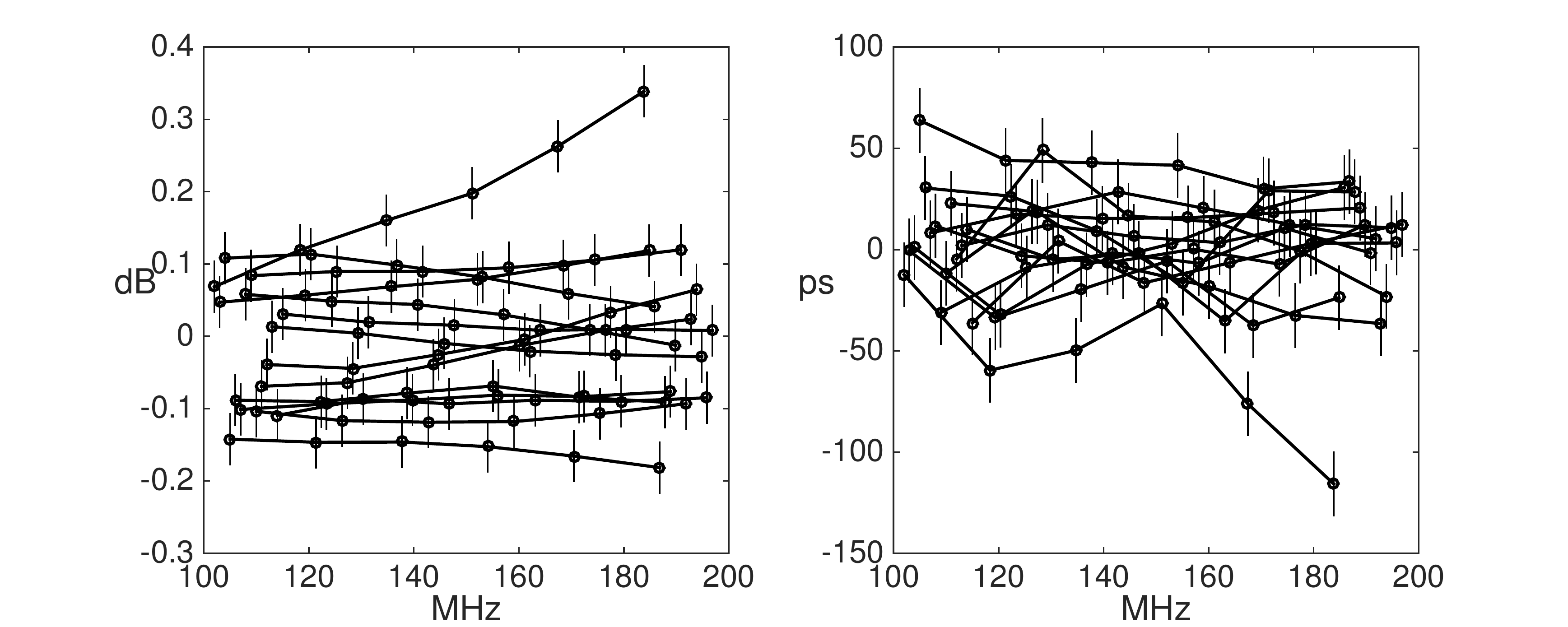}
\caption{Gain and group delay measurements on a set of 16 LNAs are shown relative to the mean LNA, as described in Sec. \ref{sec:lnameasurements}. Error bars of $\pm15$\,ps and $\pm0.035$\,dB are the RMS of repeated measuerments. At 150\,MHz, an RMS of 22ps and 0.092dB is observed. Worst cases are observed $2-3\sigma$ away from the mean.}
\label{fig:lnasplot}
\end{figure*}

\begin{figure*}[h]
\centering
\includegraphics[width=6in]{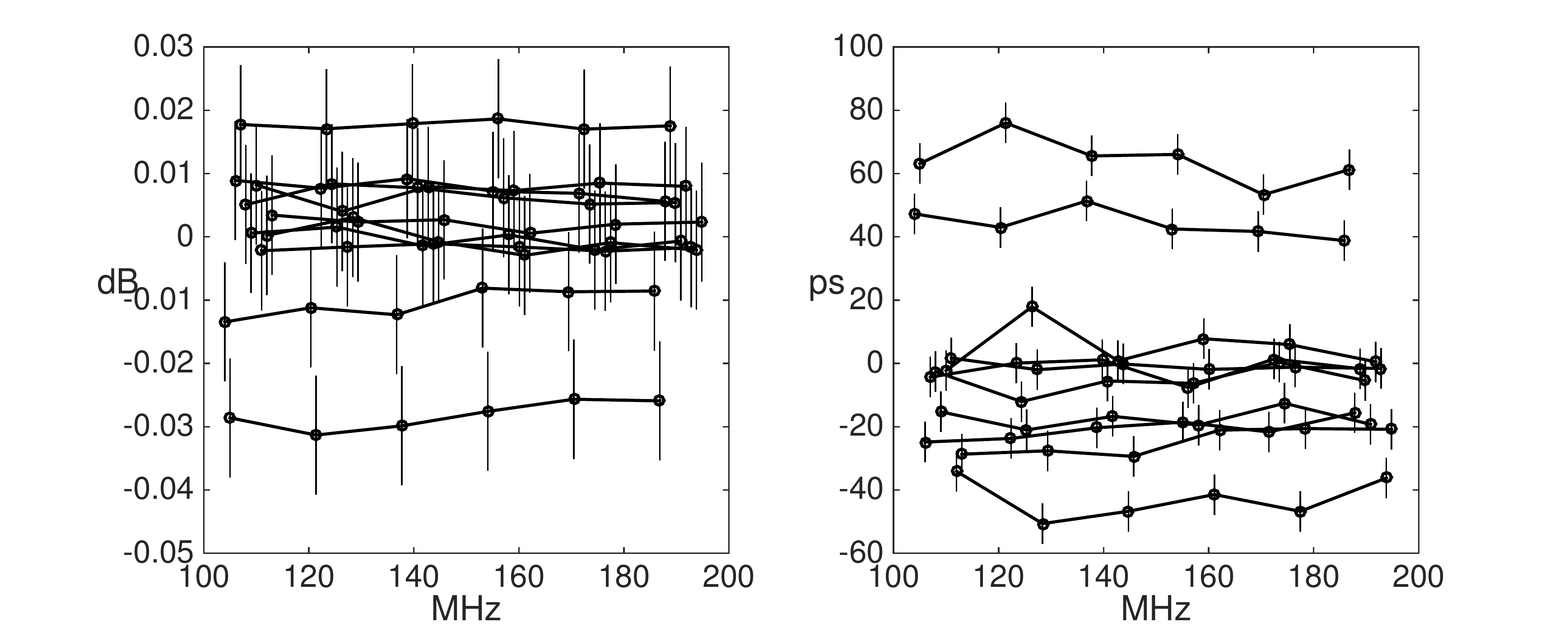}
\caption{Gain and group delay measurements on a set of 10 dipole cables are shown relative to the mean cable, as described in Sec. \ref{sec:cablemeasurements}. Error bars of $\pm6.2$\,ps and $\pm0.0093$\,dB are the RMS of repeated measurements. At 150\,MHz, an RMS of 34\,ps and 0.013\,dB is observed. Worst cases are observed $2-3\sigma$ away from the mean.}
\label{fig:cablesplot}
\end{figure*}

\begin{figure*}[h]
\centering
\includegraphics[width=6in]{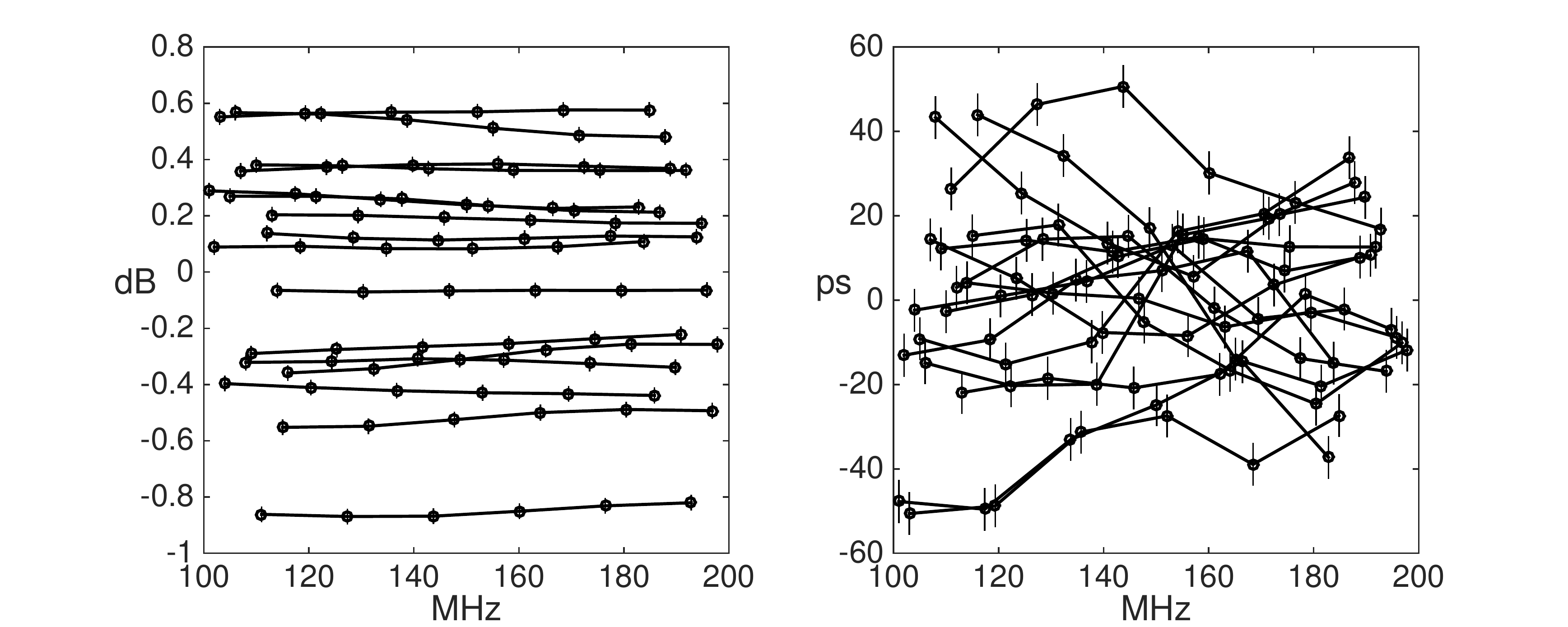}
\caption{Gain and group delay measurements on the \textit{shortest} delays of 16 beamformer inputs for one polarization are shown relative to the mean, as described in Sec. \ref{sec:bfmeasurements}. Error bars of $\pm4.9$ps and $\pm0.026$\,dB are estimated from repeatability studies. At 150\,MHz, an RMS of 21\,ps and 0.41\,dB is observed. Worst cases are observed $2-3\sigma$ away from the mean.}
\label{fig:bf00000plot}
\end{figure*}


We characterize gain and group delay variation among the cables, LNAs, and beamformer signal paths that comprise an MWA tile through precision vector network analyzer (VNA) measurements of these components. We employ an experimental setup that mitigates the challenges generally faced by such low frequency RF measurements such as reflections at interfaces or due to cable bending, parasitic RF coupling, VNA noise, and saturation of analog components. We discuss uncertainty estimation and perform repeatability checks.

In addition, tilts and misalignments of the deployed MWA tiles contribute to antenna-to-antenna beam variation and concomitant beam mismodeling. We characterize these effects using the known MWA tile positions and elevations.

\subsection{Gain and Group Delay Experiments}

Gain and group delay measurements are conducted on LNAs, dipole cables, and beamformer paths using the setups described in more detail in the sections below. In all cases, we perform measurements over the band $100-200$\,MHz, then retain the group delay and gain RMS at 150\,MHz for our beamforming error budget; the RMS is observed to be relatively frequency-independent over this band. Note that the physical gains and phases show some frequency dependence across this band, but \textit{only relative differences between the sixteen dipole pathways distort the beam pattern}. The mean gain and group delay through the 16 signal paths are absorbed into each tile's calibration amplitude and phase. For the same reason, gains and group delays of the VNA and measurement cables are irrelevant.

We use an Anritsu MS2024A vector network analyzer set to low probe power (-30\,dBm) and 30 trace averaging. The VNA is optimized for wide band (GHz) measurements, and we mitigate small-scale (sub MHz) systematics through binning to 16\,MHz. In each of these windows, we average the gains measured at 0.36\,MHz resolution, and compute the mean group delay by fitting a ramp to the measured phases. We perform repeated measurements on each component after disconnecting and reconnecting the entire measurement setup in order to estimate uncertainties due to slight bending of probe cables or imperfect cable connections.

\subsubsection{LNA Measurements}
\label{sec:lnameasurements}

Precision LNA measurements are particularly challenging in a laboratory setting given their exposed leads which, in a deployment environment, are fed balanced input by two dipole arms. Figure \ref{fig:experimentalsetup}(a) shows a diagram of our solution. We use a 180$^\circ$ two-way power splitter (Mini-Circuits ZFSCJ-2-3-S+) to split the VNA probe signal into two balanced inputs to the LNA, both mounted above an aluminum plate to mitigate RF coupling (Figure \ref{fig:newlnasetup}). The aluminum plate is grounded  to the splitter case, and then to the VNA probe cable shield. We fabricated angle connectors to secure the LNA leads to the center conductors of the power splitter outputs with as little exposed wire as possible. The LNA is powered through a Bias-T (Mini-Circuits ZFBT-4R2G-FT+) with a 5\,VDC power supply. 

We use this testing setup to characterize 16 single-polarization LNAs. Due to their different cable lead lengths, the X and Y boards have systematically different group delays which we correct for the subsequent analysis. As bending of these leads contributes to group delay variation among different LNAs, the LNA design was subsequently modified to fit both polarizations  on the same circuit board and eliminate the excess lead cable. To approximate the level of group delay variation in these dual-polarization LNAs, we estimate the group delay variance contributed by the cable leads as equal to the measurement uncertainty (assumed dominated by cable lead bending), and subtract it from the total observed group delay variance for the single-polarization LNAs. Figure \ref{fig:lnasplot} shows our measured gains and group delays with measurement uncertainties of $\pm0.03$\,dB and $\pm15$\,ps through repeated measurements on the same set of LNAs. Measurements on different LNAs are slightly offset in frequency for ease of comparison. We observe significant (relative to measurement uncertainty) gain and group delay RMS at 150\,MHz of 21\,ps and 0.09\,dB, with worst cases $2-3\sigma$ away from the mean. Subtracting (in quadrature) the 15\,ps measurement uncertainty due to cable bending from the total delay RMS yields the intrinsic LNA delay RMS of 15\,ps.

\subsubsection{Cable Measurements}
\label{sec:cablemeasurements}

Figure \ref{fig:cablesplot} shows our dipole cable gain and group delay measurements relative to an average cable, with RMS measurement errors of $\pm0.0093$\,dB and $\pm6.2$\,ps. At 150\,MHz, we observe a significant (relative to the measurement error) group delay scatter of 34\,ps RMS and an insignificant gain scatter of 0.013dB RMS. Outliers are seen $2-3\sigma$ away from the mean. The dipoles cables are specified to be phase matched to $\pm1-3^\circ$ over $100-200$\,MHz. This translates into a group delay RMS of $\pm19-55$\,ps, and is consistent with our measurements.

\subsubsection{Beamformer Measurements}
\label{sec:bfmeasurements}

Gains and group delays of a set of 16 beamformer inputs for one polarization were measured in a testing setup depicted in Figure \ref{fig:experimentalsetup}(b). To avoid bending of the VNA probe cable when moving it across the 16 beamformer inputs, a dipole cable was used to connect the VNA probe cable to the desired beamformer input. Figure \ref{fig:bf00000plot} shows our measured gains and group delays for the shortest delays on these 16 beamformer inputs with measurement uncertainties of 4.9\,ps and 0.026\,dB. We observe an RMS of 21\,ps and 0.4\,dB at 150\,MHz, with worst cases $2-3\sigma$ from the mean. The longest delays through these beamformer inputs correspond to all delay lines (``bits'') engaged, yielding $\sim13.5$\,ns of delay. We also probe the maximum delays through these beamformer inputs and find RMS's of 54\,ps and 0.43\,dB at 150\,MHz.

\subsection{Tile Tilts and Rotation}

\begin{figure*}[t]
\centering
\includegraphics[width=7in]{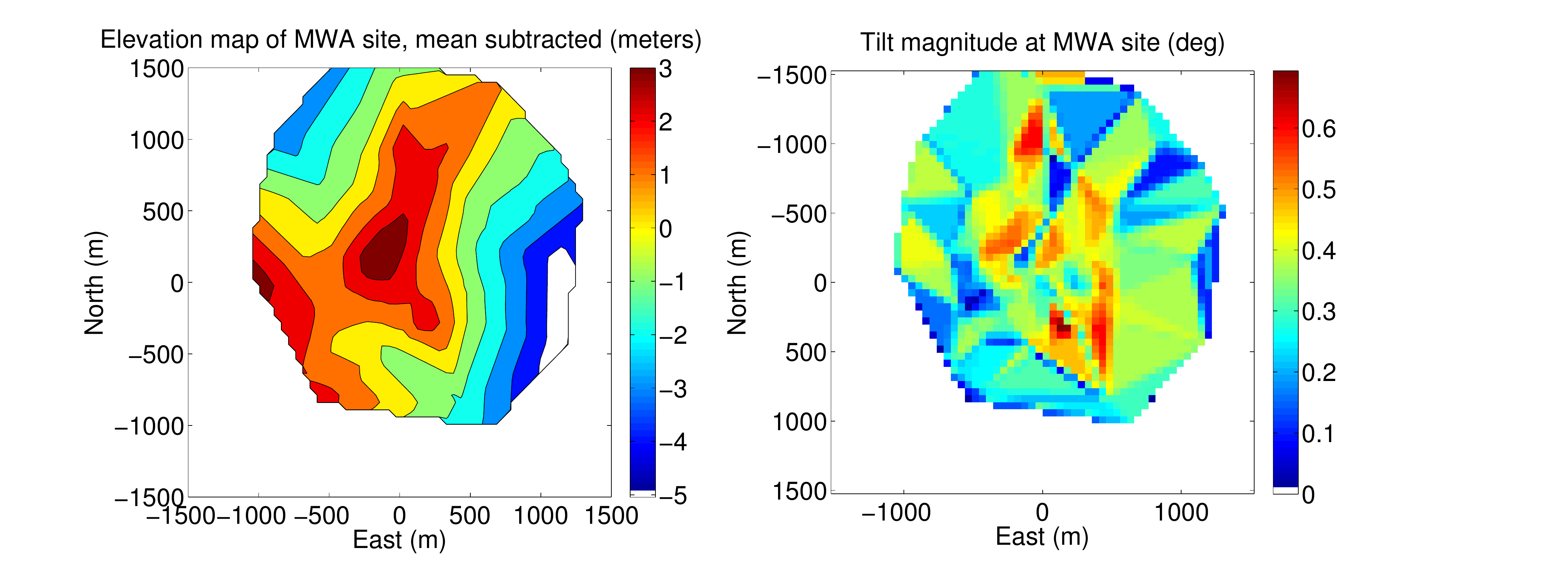}
\caption{Map of the tilt magnitude of the MWA site computed by gridding the 3D tile positions and taking the gradient. Triangular features are artifacts from sparse grid coverage by the antenna positions, nonetheless their magnitudes are likely reasonable approximations, perhaps even underestimates of the land tilts given that small scale topographic structure is unconstrained.}
\label{fig:mwatiletiltmap}
\end{figure*}

As the MWA was constructed around the apex of a slight hill to avoid flooding, a planar fit to the tile positions is quite poor. In principle tile tilts and rotations could be measured and incorporated into data reduction, however this has not yet been done. In this paper, we conservatively incorporate them into our budget of antenna-to-antenna variation. We estimate tile tilts by gridding the differential GPS mapped tile positions, then compute the magnitude of the gradient. Using a 60\,m grid spacing we find the RMS of the tilt (away from zenith) magnitude to be $0.27^\circ$, with some tiles having tilts up to $0.4^\circ$ (Figure \ref{fig:mwatiletiltmap}). These numbers are of order the precision of the differential GPS measurements used to determine the tile corners, so for simplicity we assume an RMS of $\sim0.3^\circ$ for EW tilt, NS tilt, and rotation in subsequent simulations. 

\subsection{Budget of Beamforming Errors}
\label{sec:budget}

We compile the measurements presented in this section into a budget of beamforming errors in Table \ref{table:systematictable}. Additionally we include the estimated dipole position precision of $0-17$\,ps estimated in Sec. \ref{sec:groundscreen} as it is comparable with the other sources of group delay scatter. Summing these group delay scatters in quadrature gives a total RMS of 46\,ps (68 \,ps) using the shortest (longest) beamformer delays. In contrast, the gain scatter is dominated by variation over the beamformer inputs of 0.4\,dB for both delay settings. Lastly, overall tilts and rotations of the tile with $0.3^\circ$ RMS are included separately.

 \begin{table}[h]
 \centering
 \caption{ \label{table:systematictable}Beamforming error budget at 150\,MHz.}
 \begin{tabular}{llllllll}
 \hline\hline
 Systematic Name & RMS \\
 \hline\hline
Cable group delay & 34\,ps \\
LNA delay & 15\,ps  \\
Beamformer delay (shortest delay) & 21\,ps   \\
Beamformer delay (longest delay) & 54\,ps   \\
Dipole position & $0-17$ps & \\
\hline
Cable gain & 0.013\,dB\\
LNA gain & 0.09\,dB \\
Beamformer gain (shortest delay) & 0.41\,dB \\
Beamformer gain (longest delay) & 0.43\,dB \\
\hline
Tile tilt/rotation & 0.27$^\circ$ \\
     \tableline\\
 \end{tabular}
 \end{table}

\section{Simulating Beams with Beamforming Errors}
\label{sec:simulations}

\begin{figure*}[h]
\centering
\includegraphics[width=5in]{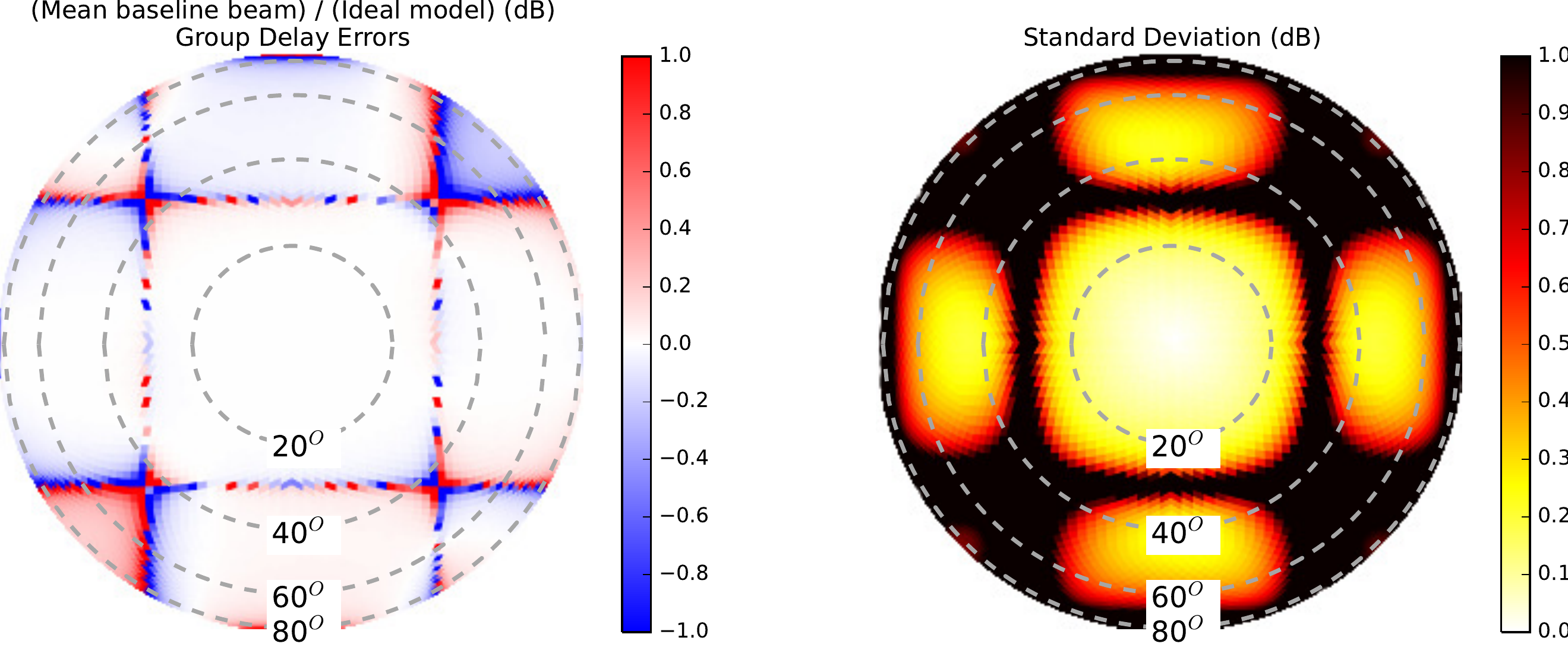}
\includegraphics[width=5in]{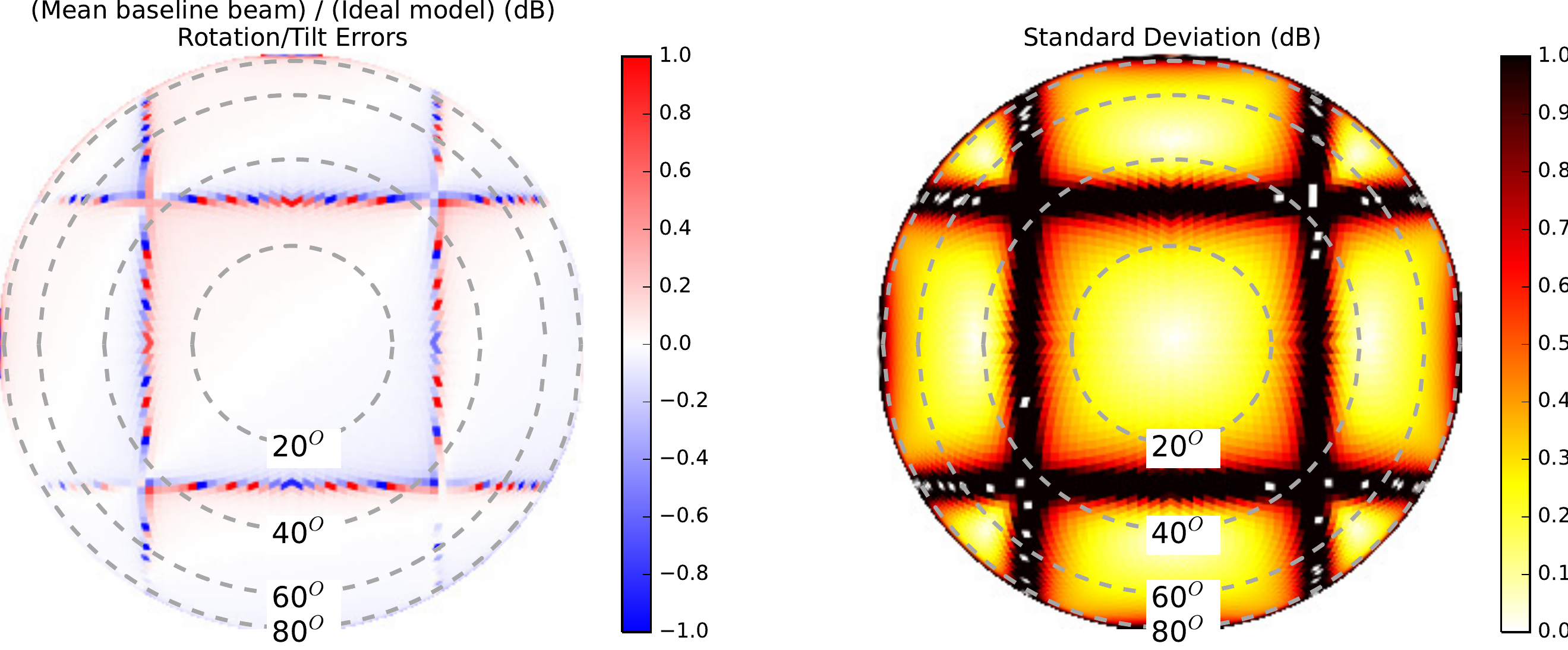}
\includegraphics[width=5in]{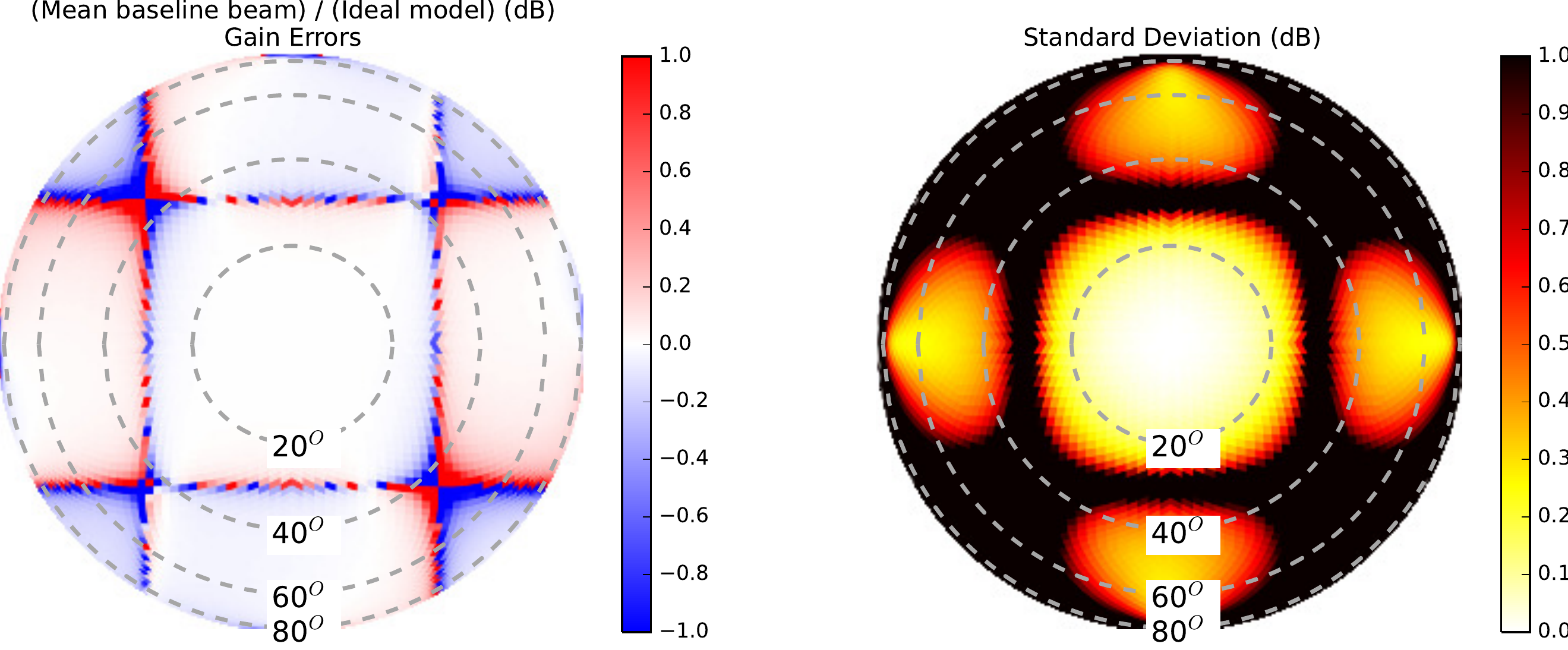}
\caption{Baseline-averaged beam (left) and standard deviation (right) of simulated beams relative to the ideal model: $\sigma_\text{delay}=50$\,ps group delays (top), $\sigma_\text{tilt,rot}=0.3^\circ$ (middle), and $\sigma_\text{gain}=0.5$\,dB (bottom). Even though the individual beams exhibit fluctuations at the $0.2-0.5$\,dB level near the edge of the mean lobe and in the sidelobes, the effects on the baseline-averaged beam are at the sub-percent level except within several degrees of the sidelobes. This is due to partial cancellation of the complex beam errors when combining the complex pair-product beams of different visibilities, here calculated assuming natural weighting. The color scale in the right panel is saturated at 1\,dB.}
\label{fig:simdelaysthetagains}
\end{figure*}

\begin{figure*}[t]
\centering
\includegraphics[width=6in]{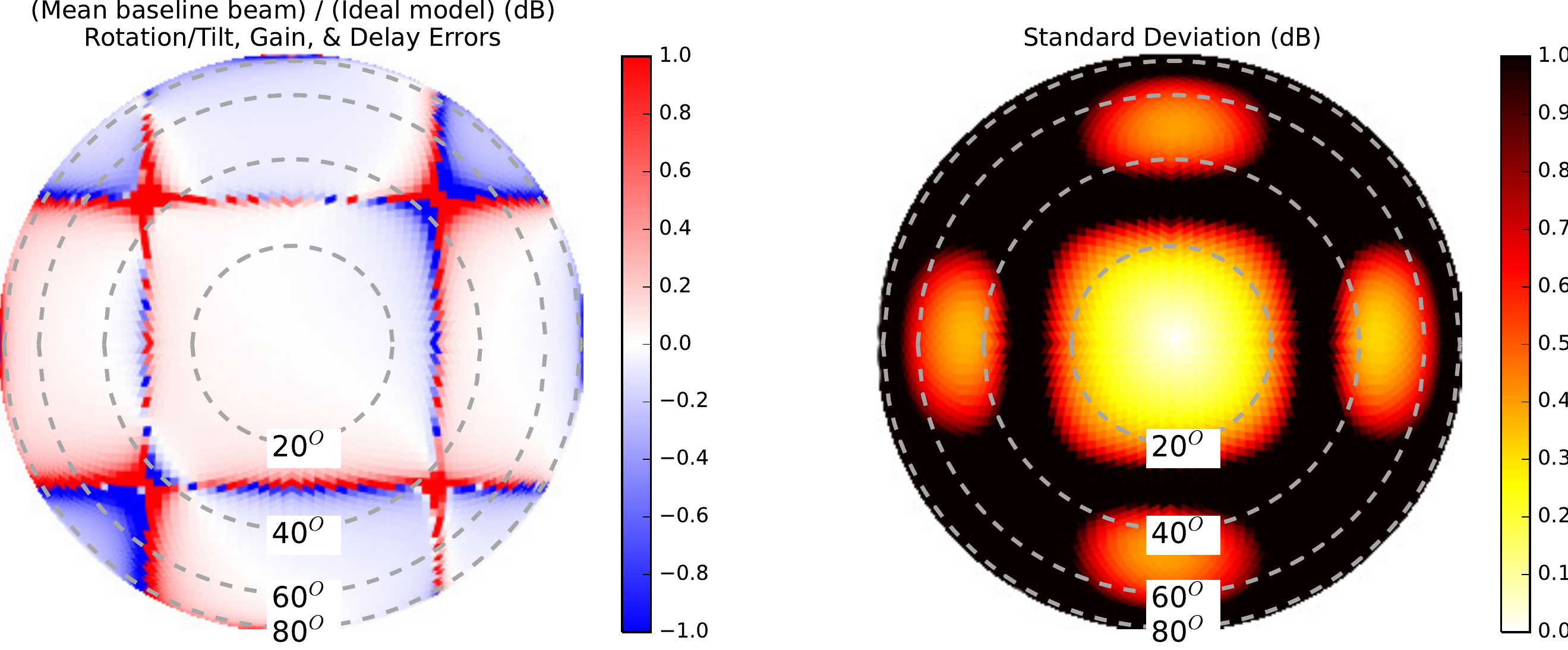}
\includegraphics[width=6in]{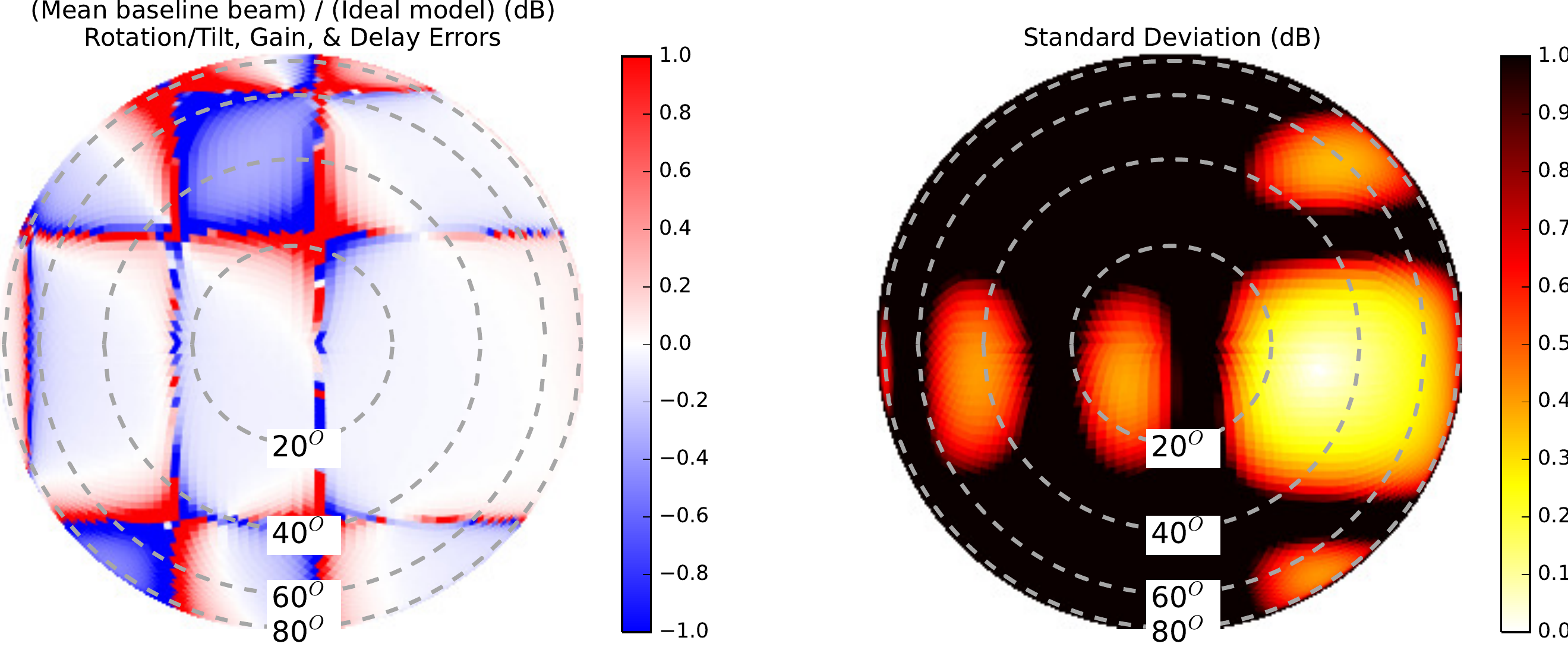}
\caption{Baseline-averaged beam (left) and standard deviation (right) of simulated beams using the full beamforming error budget ($\sigma_\text{delay}=50$\,ps group delays, $\sigma_\text{tilt,rot}=0.3^\circ$, and $\sigma_\text{gain}=0.5$\,dB) for a zenith pointing (top) and the off-zenith pointing (bottom). }
\label{fig:simall}
\end{figure*}

\begin{figure*}[h]
\centering
\includegraphics[width=5in]{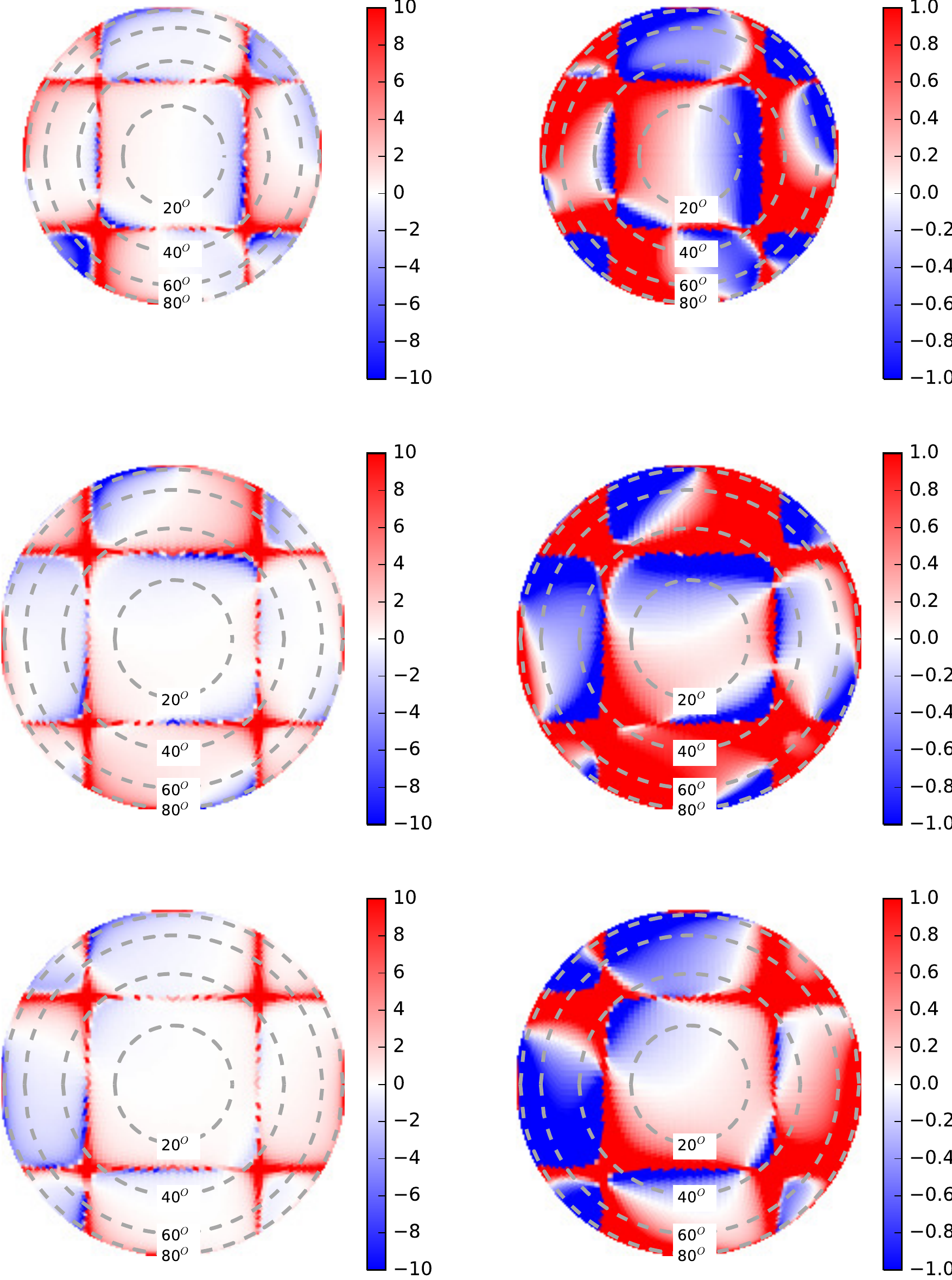}
\caption{Each row shows a realization of a simulated beam relative to the ideal beam (in dB) on a compressed color scale (left) and on an expanded color scale (right). This simulation used the full beamforming error budget of $\sigma_\text{delay}=50$\,ps, $\sigma_\text{gain}=0.5$\,dB, $\sigma_\text{tilt,rot}=0.3^\circ$. Here we see up and down fluctuations in the sidelobes and near the nulls (right) at the $\pm0.5$\,dB level seen in Figure \ref{fig:simall}, in addition to a positive bias within several degrees of the nulls (left). }
\label{fig:simallsamples}
\end{figure*}

\begin{figure*}[h]
\centering
\includegraphics[width=2.9in]{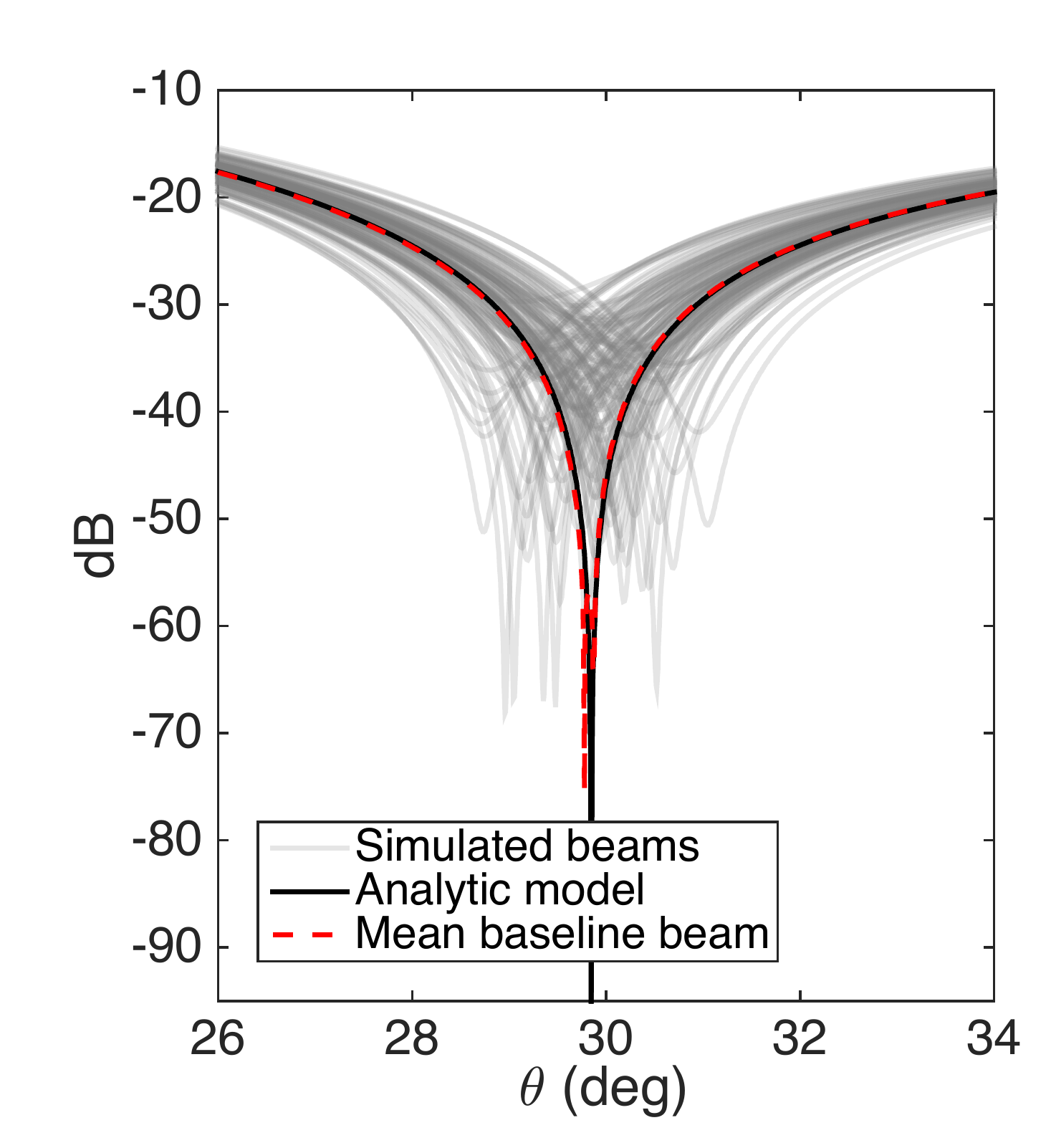}
\includegraphics[width=2.7in]{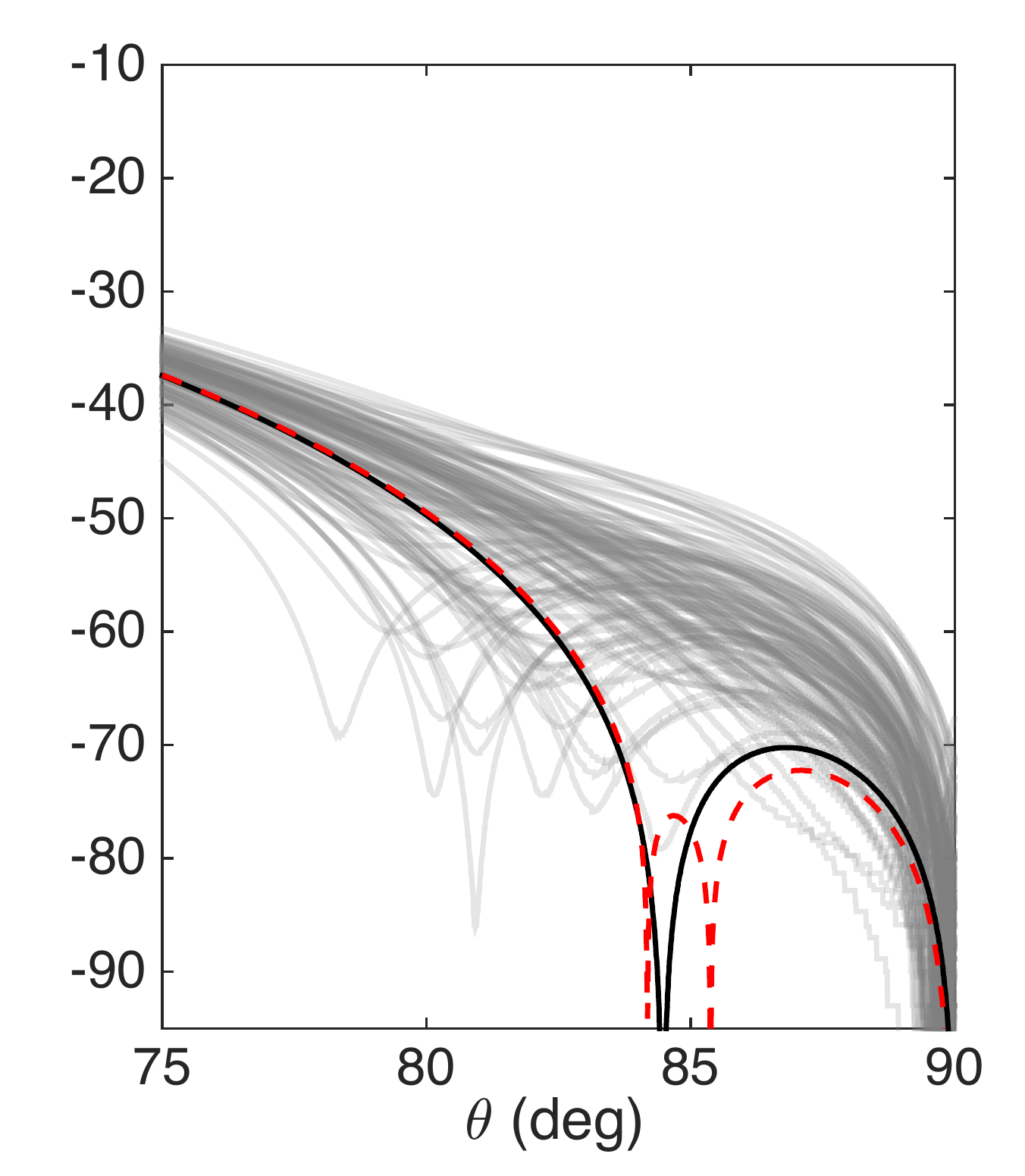}
\caption{We zoom in around the first null and near the horizon along a NS slice through the beam after running simulations with our full beamforming errors budget with $0.01^\circ$ resolution. The nulls in all 128 beams with beamforming errors ($|b_i|^2$) (gray) are ``filled in'' by the errors, however the baseline-averaged beam (Eqn. \ref{eqn:meanbaselinebeam}) (red dashed) remains very close to the ideal power beam ($|b|^2$) (black) for the reasons discussed in Sec. \ref{sec:simulations}. This demonstrates that beamforming fluctuations of different antennas tend to average out in imaging. Still, the antenna-to-antenna variation will limit deconvolution accuracy.}
\label{fig:sidelobeaveraging}
\end{figure*}

We study the separate and cumulative effects of beamforming errors on beam patterns through simulations using the budget presented in Sec. \ref{sec:budget}, assuming that the dipole gain and delay errors and the tile tilts and rotations are randomly scattered around zero. We incorporate these errors into a simple analytic beam model and compute statistics on the set of slightly corrupted beams. Extensive numerical modeling \citep{sutinjo2014} shows slight corrections relative to the analytic model towards the edge of the main lobe and in the sidelobes, especially at higher frequencies towards 200\,MHz, but is susceptible to beamforming errors in largely the same way as the analytic beam. 

The analytic electric field beam, $b(\theta,\phi,\lambda)$, models the tile as a $4\times4$ grid of EW-oriented Hertzian dipoles above a perfect, infinite ground plane, with no mutual coupling,
\begin{eqnarray}
\label{eqn:analyticbeam}
b(\theta,\phi,\lambda)=(1-e^{4\pi i h\cos\theta\lambda})\sqrt{1-\sin^2\theta\sin^2\phi}\nonumber\\
\times A(\theta,\phi,\lambda)/b_0(\lambda)
\end{eqnarray}
Here $h=0.3$m is the dipole center height above the ground screen and division by $b_0(\lambda)$ normalizes the simulated beam to unity in the boresight direction of the ideal (no beamforming errors) beam to simulate the effect of interferometric calibration. The power beam is given by $B(\theta,\phi,\lambda)=|b(\theta,\phi,\lambda)|^2$. $A(\theta,\phi,\lambda)$ is the array factor given gain errors $\{\delta G_i\}$ (dB), delay errors $\{\delta\tau_i\}$, and pointing delays $\{\tau_i\}$, defined as
\begin{equation}
A(\theta,\phi,\lambda)=\sum_{i=1}^{16}10^{\delta G_i/20}\exp(i \vec{k}\cdot \vec{x}_i-2\pi i f(\tau_i+\delta\tau_i)).
\end{equation}
where $\vec{k}$ is the wavevector of the incoming radiation. During simulations in which the tile tilt and rotation are allowed to vary, horizontal coordinates $\theta$ (zenith angle) and $\phi$ (azimuth, starting from the North, increasing towards the East) are replaced with coordinates from a tilted/rotated coordinate system.

For each of several possible systematics ($\sigma_\text{delay}=50$\,ps group delay errors, $\sigma_\text{tilt,rot}=0.3^\circ$ tile tilt/rotation errors, and $\sigma_\text{gain}=0.5$\,dB gain errors), we generate 128 tile realizations to represent the range of antenna beams in the MWA. We use the HEALPix pixelization of the sky \citep{healpix} with nside=32, corresponding to a resolution of $1.8^\circ$. This resolution is sufficient to resolve structure in the smooth beam pattern except within several degrees of the nulls. The effect of the ensemble of these slightly corrupted beams on science results depends on the type of analysis employed. In Sec. \ref{sec:effectsonpowerspectra} we consider the effects on power spectrum analyses, but we focus in this section on the effects on radio interferometric imaging. The effective beam of a naturally weighted image is the baseline-averaged beam, 
\begin{equation}
\label{eqn:meanbaselinebeam}
B_\text{baseline-averaged}(\theta,\phi)=\frac{1}{N_\text{baselines}}\sum_{i\ne j} b_i(\theta,\phi)b_j^*(\theta,\phi)
\end{equation}
Note that while the voltage beam is in general complex, the baseline-averaged beam is real because both $b_ib_j^*$ and $b_jb_i^*$ are included in the sum.

We plot in Figure \ref{fig:simdelaysthetagains} (left panel) the baseline-averaged beam relative to the ideal model for each systematic separately (delay, gain, and tilt/rotation errors), observing deviations only at the sub-percent level in the main lobe and sidelobes (though larger deviations are present within several degrees of the nulls). In the limit of infinitely many antennas these deviations would approach zero; but with only 128 antennas, these plots give a sense of the MWA's baseline-averaged beam. Beware that these sub-percent errors mask the fact that antenna-to-antenna variation will limit the accuracy of source deconvolution. To quantify the level of antenna-to-antenna variation implied by our beamforming error budget, we plot in the same figure (right panel) the standard deviation of the ratio of beam power response in each sky pixel to ideal beam power over the set of 128 simulated beams. The standard deviation is computed over this set of beam ratios in dB. We observe that individual beam realizations exhibit fluctuations at the level of $0.2-0.5$\,dB towards the edge of the main lobe ($\theta\gtrsim20^\circ$) and in the sidelobes with the tilt/rotation errors producing the smallest effects. The effects of the gain and delay errors appear similar in magnitude, and all exhibit large fluctuations near nulls where our dB standard deviation metric ceases to be meaningful.

Next we simulate a beam with the entire realistic systematic budget ($\sigma_\text{delay}=50$\,ps, $\sigma_\text{gain}=0.5$\,dB, and $\sigma_\text{tilt,rot}=0.3^\circ$) for both the zenith pointing and the off-zenith pointing  (Figure \ref{fig:simall}). In aggregate, these errors manifest as fluctuations at the level of 0.5\,dB near the edge of the mean lobe $\theta\sim20^\circ$, and $0.5-0.75$\,dB ($10-20\%$) in the sidelobes, as seen in the standard deviation plots. We also plot three sample realizations (Figure \ref{fig:simallsamples}) of these corrupted beams relative to the ideal ones, which clarify the effects on individual tile beams. These realizations also illustrate the improvements which could be achieved through use of per-antenna complex primary beams in the analysis. The left column shows these beams relative to the ideal model on a compressed color scale highlighting the effects on the nulls. The bias within several degrees of the nulls is at the $\sim5-10$\,dB level, though these exact numbers depend somewhat on pixel size as the beam is changing rapidly in these regions. Note that despite this consistent power bias, the random beam phases in these regions produce a baseline-averaged beam without such a bias (Fig. \ref{fig:simdelaysthetagains}, \ref{fig:simall}). The right column shows these same ratio plots but with expanded color scales highlighting the $0.5-1$\,dB level fluctuations seen in the main lobe and in sidelobes, a factor of a few larger than those observed in Fig. \ref{fig:simall} for each systematic individually. These fluctuations are unsurprisingly asymmetric, but appear coherent on the scale of a sidelobe. 

Lastly, we consider in more detail the effects of beamforming errors on the nulls near the main lobe and near the horizon. We first rerun our simulations with finer angular resolution of $0.01^\circ$ on a slice through the NS plane. We show the results in Figure \ref{fig:sidelobeaveraging} where we zoom in around the first null and near the horizon. We plot the ideal beam in black and our 128 realizations of beams with beamforming errors in gray (both plotted as power beams), noting that in both regions the beamforming errors ``fill in'' the analytically zero nulls and their surroundings so the first null bottoms out between roughly -55 and -30 dB, and the null at $\sim85^\circ$ bottoms out between -70dB and -40dB, vanishing entirely from some realizations. However, cancellation of the complex errors in the simulated beams results in a baseline-averaged beam (Eqn.\ref{eqn:meanbaselinebeam}) which tracks much more closely with the ideal beam than any individual realization. This same effect is seen in the previous figures. The deviations of the baseline-averaged beam away from the ideal beam are much smaller than those of individual antenna beams. The reason is that the baseline-averaged beam amounts to an average of $\text{Re}( b_i(\theta,\phi)b_j^*(\theta,\phi))$ over antennas $i<j$, and this can be negative near the nulls depending on gain and delay errors.

\section{Effects on Power Spectrum Analyses}
\label{sec:effectsonpowerspectra}

We present in this section a discussion and preliminary modeling of the effects of unmodeled primary beam variation among antenna elements in a 21\,cm EOR power spectrum analysis. We focus on the effects for the MWA, but comment on other power spectrum analyses as well. A comprehensive quantitative evaluation of the effects of these errors in real analysis pipelines demands detailed instrument simulations, building on those of \citet{nithya15} to take into account primary beam variation as in \citet{shaw15,asad15}. We leave this for future work, and consider here the qualitative effects of primary beam variation in interferometric calibration, forming of image cubes, and power spectrum analysis. We supplement this qualitative discussion of the effects on a power spectrum analysis using the simple delay spectrum technique of \citet{parsons12a,parsons12b} on a representative baseline.

\subsection{Calibration and Forming Image Cubes}
\label{sec:cal}
The MWA uses a sky model-based calibration scheme in which model visibilities are computed for a model sky distribution, and matched to the measured visibilities by fitting for antenna-based complex gains. Due to primary beam variation among the antenna elements, sources will appear with different amplitudes to different antennas, effectively adding a ``noise'' to measured visibilities relative to the ideal model. This noise adds to the inherent inaccuracy of the sky model. Given that such inaccuracies likely manifest most strongly from sources in the sidelobes where they are most difficult to measure, the resulting visibility errors rotate rapidly with time and frequency, suggesting time and frequency averaging of calibration solutions as a method to mitigate such sky-modeling errors \citep{braun2013}. This approach will also mitigate sky-modeling errors due to primary beam variation, though it is unknown if time and frequency averaging alone will be sufficient to isolate foregrounds away from the 21\,cm signal. \citet{dillonneben} and Beardsley et al. (in prep) assume that all MWA antennas have the same bandpass up to low order polynomial corrections, further reducing both sky modeling errors and thermal noise. In any case the more immediate cause of calibration-induced frequency structure is miscalibration of long baselines which imprints frequency structure on the sky, and thus, on short baselines, leaking power beyond their horizon limits. More detailed studies and end-to-end instrument simulations are needed to quantify the effects of calibration errors on 21\,cm analyses.

It is worth pointing out that while redundant calibration \citep{liu10,zheng14} has the advantage of being insensitive to sky modeling errors, it remains sensitive to primary beam variation which disturbs the assumption that nominally redundant baselines actually see the same sky signal. In the same manner as discussed above, time and frequency averaging of calibration solutions will help mitigate these errors here, though further study is needed to quantify these effects.

In forming image cubes from interferometric visibilities, primary beam models are used to weight different observations, form Stokes I, and perform primary beam correction \citep{williamsimaging, X13, dillonneben,ord2010,bernardi2013}. Antenna-to-antenna beam variation will disturb all these weighting steps, slightly upsetting the minimum-noise optimal weighting. Further, as noted in Sec. \ref{sec:simulations}, though the mean imaging beam, and thus, the dirty apparent source fluxes, are nearly unaffected by the beamforming errors due to cancellation of complex beam errors, the antenna-to-antenna variation will still limit deconvolution accuracy, and thus, foreground modeling accuracy. This is because beamforming errors alter the apparent point spread function because sidelobes from different visibilities now have slightly different weights which do not cancel out as they do at the exact source position. Further studies are needed to assess the effect in more detail, and quantify the deconvolution residuals.

\subsection{Power Spectrum}

\begin{figure*}[t]
\includegraphics[width=7in]{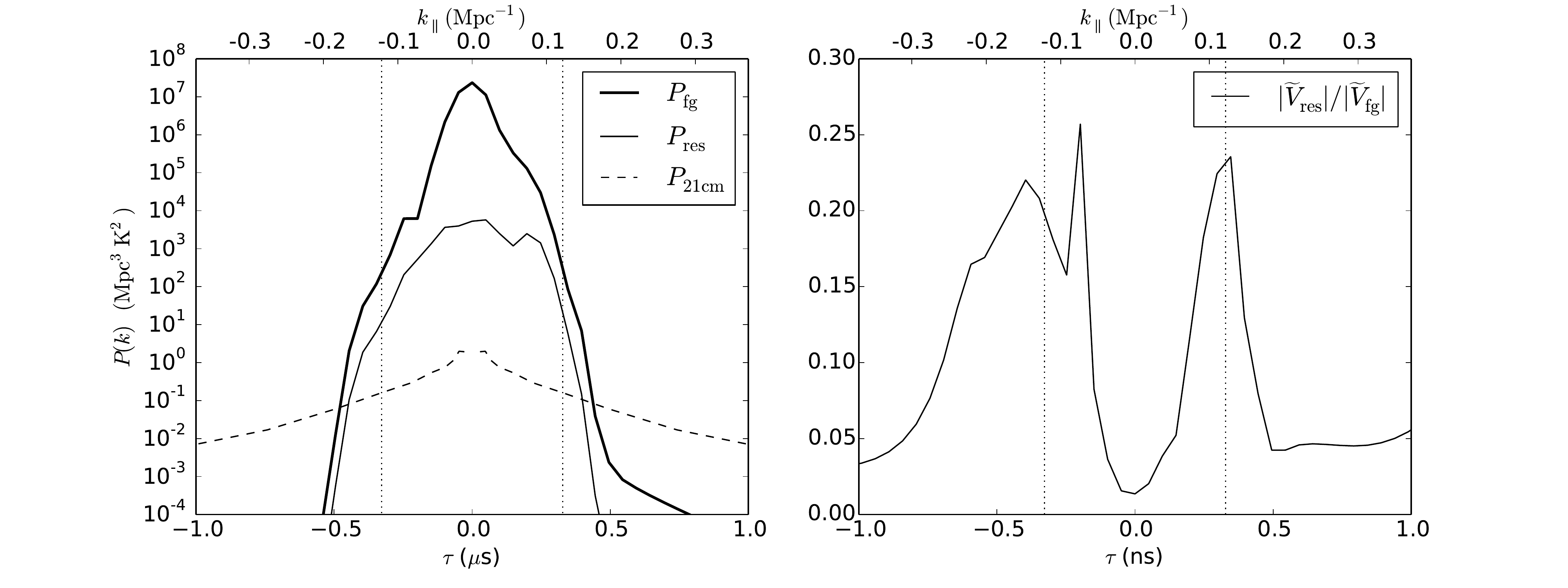}
\caption{We simulate delay power spectra for a single baseline at 150\,MHz ($z\sim8.5$) using the Global Sky Model and point source catalogs with and without beamforming errors (dipole gain and delay errors of RMS 0.5\,dB and 50\,ps and delay slope errors of RMS 5ps/\,MHz). We use a bandwidth of 20\,MHz and a Blackman-Harris window function. Left: The total foreground power $P_\mathrm{fg}$ (thick black line) is predominantly contained within this baseline's horizon limits (vertical dotted lines) where it dominates over the cosmological signal, but falls rapidly below that signal just outside the horizon limits (the ``EOR window''). This demonstrates that the ``foreground avoidance'' approach reveals the cosmological signal even in the presence of frequency dependent beamforming errors. Measurement of the cosmological signal within the baseline's horizon limits, where it is largest, requires model subtraction with $3-4$ orders of magnitude more dynamic range in power than is achieved by subtracting an otherwise perfect foreground model with unmodeled beamforming errors $P_\mathrm{res}$ (thin black line). Right: the fractional visibility residual $|V_\mathrm{res}/V_\mathrm{fg}|$ after subtraction is largest near the baseline's horizon limits (corresponding to large zenith angles near the horizon, where the effects of beamforming errors are largest, and lowest at zero delay (in the plane bisecting the baseline and including zenith). }
\label{fig:bferrorsinps}
\end{figure*}

A power spectrum analysis diverges from an imaging analysis by incoherently averaging fourier modes (to bin different $\vec{k}$ modes into a 1D power spectrum) instead of coherently averaging them. Thus even though the effects of beamforming errors on the baseline-averaged beam are small as discussed above, further operating on the image to produce a power spectrum can make these errors very significant. Even assuming all antennas are perfectly calibrated despite the primary beam variation, sources will appear slightly brighter to some antennas and slightly dimmer to others. Subtraction of a foreground model which neglects this effect by assuming ideal beam patterns leaves residuals which vary over this spherical shell, and do not average down in the incoherent power average. 

It is these effects, as opposed to calibration effects, which we expect to be the most significant in power spectrum analysis, and to further quantify them, we simulate a power spectrum on a single baseline with and without unmodeled beamforming errors. In essence, we ask what errors would we make in a power spectrum analysis if we knew the foregrounds perfectly but lacked measurements of the exact antenna-to-antenna variation. We consider their implications for two different foreground mitigation strategies: foreground subtraction and foreground avoidance. Our simulations are centered on the MWA ``EOR0'' deep integration field, centered at R.A.(J2000) $= 0^\text{h}\,0^\text{m}\,0^\text{s}$ and decl.(J2000) $= -30^\circ\,0'\,0''$, and the sky is modeled as the sum of a deep MWA point source survey within 20$^\circ$ of the field center (Carroll et al., in prep.), the shallower but wider MWA commissioning point source survey\citep{MWACS}, the Culgoora catalog\citep{Slee1995}, and the Global Sky Model \citep{gsm}. We simulate the visibilities for a 100\,m baseline measured over a 20\,MHz bandwidth centered at 150\,MHz ($z\sim8.5$), divided into 200\,kHz channels. We do this first  assuming both both antennas in the baseline are have independent beamforming errors, and then, for an ideal beam without errors. We use beamforming errors motivated by our measurements in Figures \ref{fig:lnasplot} and \ref{fig:bf00000plot}, namely dipole gain and delay errors of RMS 0.5\,dB and 50\,ps and delay slope errors of RMS 5\,ps/MHz. As the group delay frequency-dependence is not well measured on MHz frequency scales due to our group delay window size, we intend this level of delay slope RMS as a significant overestimate of the observed delay slopes in our measurements. It is meant to set an upper limit on the effect of frequency dependent beamforming errors on the critical frequency dimension of 21\,cm measurements. We neglect tile tilt/rotation errors as Sec. \ref{sec:simulations} suggests they are subdominant to gain and delay errors. 

We plot in the left panel of Figure \ref{fig:bferrorsinps} the mean foreground power spectrum computed from 100 realizations of simulated visibilities with beamforming errors, $P_\mathrm{fg}$, and then the power spectrum after subtracting ideal model visibilities, $P_\mathrm{res}$. The delay power spectrum is computed as outlined by \citet{nithya15} using a Blackman-Harris Window function \citep{parsons12a,parsons12b}. The sky power (thick black line) dominates over the cosmological signal by typically $4-7$ orders of magnitude in power within this baseline's horizon limits (the well-known foreground ``wedge'') \citep{Dattapowerspec,X13, PoberWedge,MoralesPSShapes, VedanthamWedge, nithya13, CathWedge, AdrianWedge1, AdrianWedge2}, but quickly drops below the cosmological signal (dashed line) \citep{21cmfast} outside these limits (in the the ``EOR window''). This demonstrates the ``foreground avoidance'' approach and shows that it reveals the cosmological signal even in the presence of frequency dependent beamforming errors. Measurement of the cosmological signal within the baseline's horizon limits, where it is largest, requires model subtraction with $2-4$ orders of magnitude more dynamic range than is achieved by neglecting beamforming errors in the foreground model (thin black line). Note that \citet{nithya15,nithya15b} observe that an increased near-horizon beam response relative to our analytic tile model tends to add a power bump at the baseline's horizon limits (the outer prongs of their ``pitchfork'').

To be sure, these residuals due to beam errors will average down somewhat when different baselines are coherently averaged in the same $\vec{k}$ cell, but $10^4-10^8$ independent samples would be needed to bring them below the level of the EOR. The maximum number of independent samples is the number of antennas in the array, each with a different realization of beamforming errors. We thus see that the coherent averaging down of beam errors in imaging power spectra is only a small effect.

To better understand these results, we plot in the right panel of Figure \ref{fig:bferrorsinps} the residual of the simulated visibilities in delay space relative to those without beamforming errors $|\widetilde{V}_\mathrm{res}/\widetilde{V}_\mathrm{fg}|$, where $\widetilde{V}$ represents the frequency fourier transform of $V(f)$. As expected, the fractional residuals are largest ($\sim$20\%) near the delays corresponding to the baseline's horizon limit (300\,ns) as these delays correspond to very low points in the beam where the effects of the beamforming errors are largest. At zero delay, corresponding to emission from the plane bisecting the baseline vector and containing zenith, the fractional residual is much lower (1.5\%). This highlights again that beam modeling errors affect preferentially the weakest beam regions which, because they are closest to the horizon, are most at risk of leaking power into the EOR window. 

\section{Discussion}

Efforts to detect neutral hydrogen emission at cosmological distances in the presence of bright galactic and extragalactic foregrounds are drawing attention to radio astronomy systematics, in particular primary beam characterization. Following up on efforts to constrain the mean MWA tile beam through advanced modeling  \citep{sutinjo2014} and measurements \citep[][]{NebenOrbcommPaper}, we explore the next order effect of antenna-to-antenna variation. We establish a budget of relevant beamforming errors and run simulations drawing from this distribution to study the effects on beam patterns. 

We characterize the beamformer paths, dipole cables, and LNAs used in the MWA tile through laboratory experiments. Summing in quadrature the group delay errors of the cables, the LNAs, and the beamformer paths, we find 46\,ps of group delay RMS, and 67ps when using the longest beamformer paths instead. This level is roughly 10\% of the beamformer shortest delay of 435ps. Gain errors appear dominated by 0.5\,dB RMS among the beamformer inputs. These errors, in addition to tile tilt/rotation errors at the $\sim0.3^\circ$ level will vary from tile to tile yielding visibility errors which do not average down with time.

We run simulations drawing from these gain, delay, and alignment errors to study the magnitude and angular-dependence of the resulting beam errors. None of these systematics is observed to have more than a percent effect on the baseline-averaged beam (the effective beam of an image) except within several degrees of the nulls. In contrast, power spectrum measurements are more sensitive to the beam standard deviations, essentially the typical tile-to-tile variation as a function of angle on the sky. Standard deviations of roughly $0.5-0.75$\,dB ($10-20\%$) are observed towards the edges of the main lobe ($20^\circ<\theta<40^\circ$) and in the sidelobes when all systematics are included (Figure \ref{fig:simall}). 

To study the effects of these beamforming errors on 21\,cm power spectrum analyses, we break down such an analysis into the different steps where beamforming errors could affect the results, and qualitatively evaluate their likely severity. They will limit calibration fidelity, though averaging in time, frequency, and over antennas can mitigate them to some extent. While the effect on the effective imaging beam will be small due to cancellation of complex visibility errors in imaging (due to \textit{coherent} combination of fourier modes), antenna-to-antenna variation will limit deconvolution accuracy nonetheless. By the same token, the effects in the power spectrum space will be larger as here different fourier modes are added \textit{incoherently} binning fourier modes into a 2D or 1D power spectrum. 

We confirm this with a simple simulation of the delay spectrum of a single visibility, addressing the question of what errors we would make in a power spectrum analysis if we knew the foregrounds perfectly but lacked measurements of the antenna-to-antenna beam variation. We find that unmodeled beamforming errors are severe enough to make foreground subtraction impossible within the baseline's horizon limits (in the wedge), where per-antenna primary beams will be necessary. However, even including an overestimate of their frequency dependence, the beamforming errors do not leak significant frequency structure into ``the EOR window'' which remains nearly clear of foreground contamination. Thus the foreground avoidance approach being pursued by PAPER, the MWA, and HERA will remain valid even in the presence of beamforming errors.

The possibility of antenna-to-antenna variation was certainly not unexpected, though measuring beams of all 128 deployed MWA antennas remains a challenge. Improved satellite-based beam calibrators and drone-based beammapping systems are under study and may make per-antenna beam measurement a reality, capturing the additional real world effects of uneven wear and tear and even failed components. Independently, future work will extend simulations by \citet{nithya15} to include per-antenna beamforming errors and propagate them from measured visibilities through calibration, imaging, and power spectrum analysis to definitely address their effects on 21\,cm science for the MWA. 

Building on lessons learned from development of the MWA and PAPER, HERA is pursuing a targeted experiment to detect the cosmological signal using zenith-tracking dishes rather than phased arrays, and foreground avoidance rather than the more challenging subtraction. In contrast, observatories relying both on beamforming and foreground subtraction (e.g., LOFAR and SKA-Low) must model the sky and primary beams (either through calibration or measurement) to four to five orders of magnitude of dynamic range lest foreground residuals swamp the feeble cosmological signal.

\acknowledgments
This work was supported by NSF grant AST-0821321, the Marble Astrophysics Fund, and the MIT School of Science. We thank Aaron Ewall-Wice and Hamdi Mani for assistance in running these experiments, Steve Burns for debugging beamformer issues, and Danny Jacobs, Nithyanandan Thyagarajan, and Lu Feng for helpful discussions. 

This scientific work makes use of the Murchison Radio-astronomy Observatory, operated by CSIRO. We acknowledge the Wajarri Yamatji people as the traditional owners of the Observatory site. Support for the MWA comes from the U.S. National Science Foundation (grants AST-0457585, PHY-0835713, CAREER-0847753, and AST-0908884), the Australian Research Council (LIEF grants LE0775621 and LE0882938), the U.S. Air Force Office of Scientific Research (grant FA9550-0510247), and the Centre for All-sky Astrophysics (an Australian Research Council Centre of Excellence funded by grant CE110001020). Support is also provided by the Smithsonian Astrophysical Observatory, the MIT School of Science, the Raman Research Institute, the Australian National University, and the Victoria University of Wellington (via grant MED-E1799 from the New Zealand Ministry of Economic Development and an IBM Shared University Research Grant). The Australian Federal government provides additional support via the Commonwealth Scientific and Industrial Research Organisation (CSIRO), National Collaborative Research Infrastructure Strategy, Education Investment Fund, and the Australia India Strategic Research Fund, and Astronomy Australia Limited, under contract to Curtin University. We acknowledge the iVEC Petabyte Data Store, the Initiative in Innovative Computing and the CUDA Center for Excellence sponsored by NVIDIA at Harvard University, and the International Centre for Radio Astronomy Research (ICRAR), a Joint Venture of Curtin University and The University of Western Australia, funded by the Western Australian State government.

\appendix

\section{Design of the MWA Antenna Tile}

\subsection{Design and Science Requirements}
Redshifted hydrogen line emission from the Epoch of Reionization ($6\lesssim z\lesssim12$) appears in the 100-200\,MHz band, several orders of magnitude fainter than galactic and extragalactic radio emission. Separating this high redshift signal from foregrounds is thought to be possible by exploiting their different frequency dependence. While the foregrounds result from smooth spectrum radio synchrotron emission, the frequency axis of the signal is actually a cosmological redshift axis, and thus probes the complex spatial structure of the ionizing universe. Instrumental noise also plays a key role, and its mitigation necessitates large collecting area.

These science goals informed the instrumentation requirements as follows. The desired frequency band represents an order unity fractional bandwidth, necessitating a wideband antenna with a smooth frequency response. In particular, significant frequency structure on scales smaller than the nominal power spectrum analysis bandwidth of 10\,MHz (set by timescale of $\Delta z\sim0.5$ over which the cosmological signal is expected to evolve) would complicate beam modeling and risk smearing smooth spectrum foregrounds into spectrally noisy signal-like modes. A large field of view, in comparison to more traditional radio telescopes like the Very Large Array, is also desired to maximize the cosmological volume over which to measure the EOR power spectrum. Instrumental noise is minimized to the sky noise limit through use of low noise amplifiers (LNAs) placed as early in the signal chain as possible. A steerable beam was also desired to allow deep observing on discrete patches of sky, and thus coherent integration on power spectrum modes. This strategy reduces noise much more quickly with integration time at the expense of an increase in cosmic variance noise \citep{TrottObservingModes}. Lastly, and arguably most importantly, is the large required collecting area at modest cost, achieved with an array of order one hundred low cost ($\lesssim\$2500$/ea) antenna elements. The MWA is a realization of the ``Large N--Small D'' array concept consisting of a large number of small diameter antenna element made possible by advances in parallel computer processing. 

Though it is tuned to some extent to achieve the high surface brightness sensitivity required by EOR science, this design also permits a host of other low frequency science ranging from transient searches and source surveys to solar and ionospheric science \citep{mwascience}. The MWA tile design is, therefore, a compromise to meet different science goals. Further optimization for EOR science is possible, for example, HERA, is pursuing larger antenna elements to increase sensitivity without extra computing cost. At the array level, the HERA antennas will be positioned on a regular grid to achieve many redundant baselines, and thus allow coherent integration on individual power spectrum modes.

\begin{figure*}[h]
\includegraphics[width=7in]{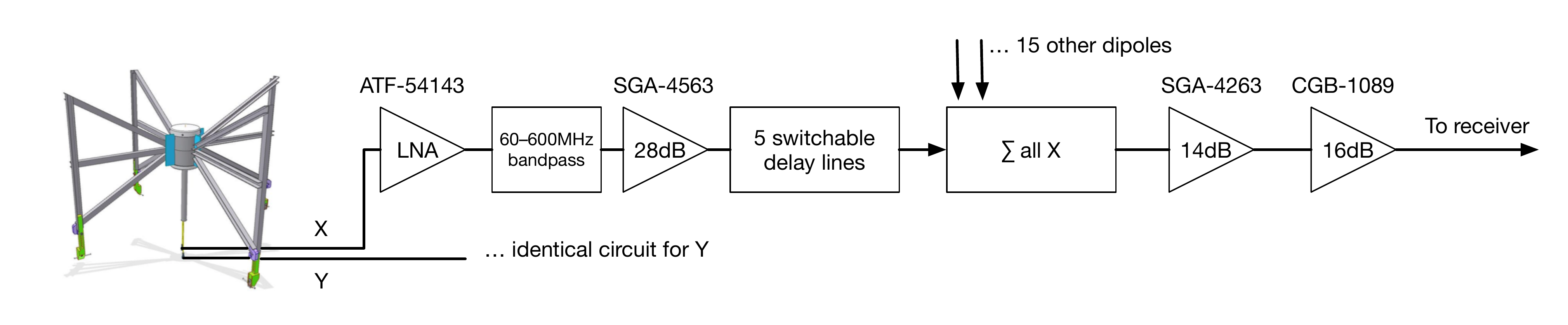}
\caption{For each polarization, each of 16 dipole signals passes through a balun/LNA (represented by the first amplifier in the diagram) mounted in the dipole hub with a gain varying between 16--25\,dB across the band. The signal is then carried into the beamformer, where it goes through a low pass filter, an amplifier, 5 switchable delay lines, a series of two-way power combiners which sums the 16 dipole signals, then two more amplifiers. The delay lines are replaced by matching attenuators when disengaged. Walsh switching may be implemented directly after the power combiners. Driving just one beamformer input, the gain totals roughly 33\,dB at 150\,MHz accounting for losses in the power combiners and other components. }
\label{fig:beamformersdiagram}
\end{figure*}

\subsection{Dipole Element}

Each MWA dipole element is a set of two orthogonally crossed vertical bowties, each of length 74\,cm and height 38\,cm. Each bowtie is composed of two aluminum arms mounted at a PVC hub such that the lowest part of the antenna is 8\,cm above the ground screen. In principle, an infinite bowtie antenna has infinitely broad bandwidth because it has no characteristic length scale. A real bowtie is truncated, which introduces length scales and resonances, but the bandwidth remains broad and the response generally smooth on frequency scales of interest. 

Note that despite the arms not being composed of solid metal sheets, the electrical performance of our bowtie differs negligibly from that of a more costly solid one while also mitigating wind loading. Other dipole style antennas were modeled in various orientations, but the bowtie was chosen for its relatively smooth gain over frequency, minimum gain variation over elevation, low horizon gain, absence of blind spots or other anomalies in the patterns or impedance, and impedance match with the LNAs.

One unexpected discovery was made during early antenna testing relating to coupling between adjacent dipoles in a tile. Interactions between the vertical pieces in a row or column of dipoles direct power towards the horizon in much the same way that one attaches perpendicular arms to a metal rod to form a directive Yagi antenna. Consequently, the dimensions of the antenna were adjusted to the present values to move this resonance to 240\,MHz, near an already unusable satellite band. 

Inside each PVC hub is a dual-polarization low-noise amplifier (LNA) that also serves as a balun between the balanced bowtie terminals and the 50\,$\Omega$ coaxial cable to the beamformer. In detail, there is an amplifier on each of the four dipole arms. The amplified signals from opposing arms are combined through a center-tapped transformer balun to feed a 50\,$\Omega$ unbalanced coaxial cable.The LNA gain with a 50\,$\Omega$ source is $\sim$19\,dB at 150\,MHz. The impedance match with the sky is sufficient to make the system sky noise dominated. Quality control data on the field-deployed dipole LNAs is collected periodically in an array dipole testing mode during which the beamformer paths are all switched off, then each is switched on individually. The LNAs are powered with a 5\,V DC bias provided by the beamformer on the 7\,m LMR-100 cables (50$\Omega$) which also carry the sky signals in the opposite direction. In their deployed configuration, these cables are fixed with wire ties atop the ground screen. 

\subsection{Ground Screen}
\label{sec:groundscreen}
The 5\,m $\times$ 5\,m ground screen is formed by three overlapped 2\,m $\times$ 5\,m mesh panels welded together and laid directly on the ground. Each panel is constructed of 3.15\,mm galvanized steel wire, welded together to form a grid of 50\,mm $\times$ 50\,mm squares. Typical dipole position errors are at the 5\,mm level or smaller on average throughout the dipole grid, due to mesh thickness and distortions due to handling, and also slightly larger errors overlapping the different mesh panels. Such horizontal errors are irrelevant for radiation incident from zenith, but contribute per-dipole delays up to an RMS of approximately 17\,ps towards the horizon due to the altered light travel time to the different elements. In any case, these errors are subdominant to other errors discussed below.

No large or small scale ground leveling was attempted, but the flatness and alignment estimated with  differential-GPS measurements is better than a few cm vertically and $\sim1^\circ$ in alignment with North and zenith. We discuss this alignment precision in more detail in Sec. \ref{sec:measurements}. An electrical path to ground is provided through a connection to the wire chassis of the beamformer and subsequently to the receiver, itself grounded to metal ground stakes.

\subsection{Beamformer}
The beamformer contains two vertically offset delayline boards, one for each polarization, each fed by 16 dipole inputs. Each input is directed through a series of digitally switchable delay lines before being summed with the others with specified relative delays applied, and output to the receiver. Figure \ref{fig:beamformersdiagram} shows a block diagram of the signal path. Each input passes through a 4-pole lowpass filter with a 3\,dB cutoff at 600\,MHz, a 30\,dB amplifier, five sequential switched delaylines, a switch that either passes the signal or terminates with 50\,$\Omega$, and lastly a cascade of two-way power combiners which sum the 16 inputs. Note that this low pass filter serves simply to protect the analog components from saturation; there is a second low pass anti-aliasing filter in the receiver. The shortest delay line is 435\,ps, a number determined by the requirement that the beam be steerable to $30^\circ$ elevation with five delay line ``bits'' whose electrical lengths form a geometric series with a ratio of 2. These boards are also capable of applying Walsh switching to the summed signal to mitigate cross-coupling between different signal paths from different MWA tiles, though this feature has not been found to be necessary and has not been implemented. 

Digital communication to the beamformer to activate delay bits on each of the delay lines is transmitted in a ``data over coax'' configuration, multiplexed on the two RG-6 cables carrying the dual-polarization beamformer output to the receiver. 




\bibliography{beamforming_errors_in_mwa_tiles}

\clearpage

\end{document}